\documentclass[a4paper,11pt]{article}
\usepackage[DIV12]{typearea}
\usepackage{longtable}
\usepackage{booktabs}
\usepackage{multirow}
\usepackage{amsmath}
\usepackage{amssymb}
\usepackage{bbm}
\usepackage[utf8]{inputenc}
\usepackage{pdflscape}
\usepackage{xspace}
\usepackage[caption=false]{subfig}
\usepackage{nicefrac}
\usepackage{graphicx}
\usepackage{cite}  
\usepackage[small]{caption2}

\newcommand{\rep}[1]{\ensuremath\boldsymbol{#1}}

\newcommand{\Z}[1]{\ensuremath{\mathbbm{Z}_{#1}}} % z_N ->\Z{N}

\newcommand{\SL}[1]{\ensuremath{\mathrm{SL}(#1)}}

\newcommand{\I}{\mathrm{i}}
\newcommand{\Id}{\mathbbm{1}}

\newcommand{\CP}{\ensuremath{\mathcal{CP}}\xspace}
\newcommand{\x}{\ensuremath{\times}}
\newcommand{\vev}[1]{\ensuremath{\langle{#1}\rangle}}

\usepackage{xcolor}
\definecolor{darkgreen}{HTML}{109930}
\definecolor{pink}{rgb}{0.858, 0.188, 0.478}

\advance \headheight by 3.0truept       % for 12pt mandatory...
\usepackage[pdftex]{hyperref}
\hypersetup{
    pdftitle = {The eclectic flavor symmetry of the Z2 orbifold},
    pdfauthor = {Baur, Kade, Nilles, Ramos-Sanchez, Vaudrevange}
}

\begin{document}

\begin{titlepage}

\begin{flushright}
\normalsize{TUM-HEP 1276/20}
\end{flushright}

\vspace*{1.0cm}

\begin{center}
{\Large\textbf{\boldmath The eclectic flavor symmetry of the \Z2 orbifold }\unboldmath}

\vspace{1cm}

\textbf{Alexander Baur$^{a,c}$, Moritz Kade$^{a}$},
\textbf{Hans Peter Nilles$^{b}$, Sa\'ul Ramos--S\'anchez$^{a,c}$, Patrick K.S. Vaudrevange$^{a}$}
\\[8mm]
\textit{$^a$\small Physik Department T75, Technische Universit\"at M\"unchen,\\ James-Franck-Stra\ss e 1, 85748 Garching, Germany}
\\[2mm]
\textit{$^b$\small Bethe Center for Theoretical Physics and Physikalisches Institut der Universit\"at Bonn,\\ Nussallee 12, 53115 Bonn, Germany}
\\[2mm]
\textit{$^c$\small Instituto de F\'isica, Universidad Nacional Aut\'onoma de M\'exico,\\ POB 20-364, Cd.Mx. 01000, M\'exico}
\end{center}

\vspace{1cm}

\vspace*{1.0cm}

\begin{abstract}
Modular symmetries naturally combine with traditional flavor symmetries and $\CP$, giving rise to 
the so-called eclectic flavor symmetry. We apply this scheme to the two-dimensional $\Z{2}$ 
orbifold, which is equipped with two modular symmetries $\SL{2,\Z{}}_T$ and $\SL{2,\Z{}}_U$ 
associated with two moduli: the K\"ahler modulus $T$ and the complex structure modulus $U$. The 
resulting finite modular group is $((S_3\times S_3)\rtimes \Z4)\x\Z2$ including mirror symmetry 
(that exchanges $T$ and $U$) and a generalized \CP-transformation. Together with the traditional 
flavor symmetry $(D_8\x D_8)/\Z2$, this leads to a huge eclectic flavor group with 4608 elements. 
At specific regions in moduli space we observe enhanced unified flavor symmetries with as many as 
1152 elements for the tetrahedral shaped orbifold and $\vev T=\vev U=\exp(\nicefrac{\pi\I}{3})$. 
This rich eclectic structure implies interesting (modular) flavor groups for particle physics 
models derived form string theory.
\end{abstract}

\end{titlepage}

\newpage

%%%%%%%%%%%%%%%%%%%%%%%%%%%%%%%%%%%%%%%%%%%%%%%%%%%%%%%%%%%%%%%%%%%%%%%%%%%%%%%%%%%%%%%%%%%%%%%%%%%%%%%%%%%%%%%%%%%%%%%%%%%%%%%%%%%%%%%%%
\section{Introduction}

In the present paper we further explore the eclectic flavor picture in the framework of orbifold 
compactifications of string theory. The eclectic flavor symmetry is the maximal discrete symmetry 
that can arise from the nontrivial combination of traditional flavor symmetry and modular flavor 
symmetry~\cite{Nilles:2020nnc}. In this approach, the benefits of finite modular symmetries, 
uncovered by ref.~\cite{Feruglio:2017spp} and further developed for example in refs.~\cite{
Novichkov:2019sqv, %1905.11970
Liu:2019khw,       %1907.01488
Kobayashi:2019xvz, %1909.05139
Liu:2020msy        %2007.13706
}, are merged with the appeal of traditional flavor symmetries (see e.g.~\cite{Feruglio:2019ktm}), 
possibly improving~\cite{Nilles:2020kgo} their predictability~\cite{Chen:2019ewa}. Up to now the 
detailed analysis has concentrated on two-dimensional orbifolded tori where the complex structure 
modulus $U$ is fixed geometrically to allow for the specific orbifold twist of the 
torus~\cite{Baur:2019kwi,Baur:2019iai}. Aspects of the embedding of the two-tori into 
six-dimensional compactified space have been discussed in ref.~\cite{Nilles:2020tdp} and have shown 
to be especially relevant for the discussion of $R$-symmetries. To capture the full eclectic 
picture one has to consider also those orbifolded tori, where the complex structure modulus is not 
fixed. This will certainly lead to a richer eclectic structure, as modular transformations now act 
nontrivially on both, the K\"ahler modulus $T$ and the complex structure modulus $U$. As a first 
step in this analysis, we consider the $\mathbbm T^2/\Z2$ orbifold that allows the full action of 
$\SL{2,\Z{}}_T$ and $\SL{2,\Z{}}_U$. This captures all the qualitative aspects of the eclectic 
flavor picture and can be used as building block for the general discussion in the six-dimensional 
case.

Flavor symmetries of the $\mathbbm T^2/\Z2$ orbifold strongly depend on the values of the moduli. 
First, we discuss in section~\ref{sec:Traditional} the case for generic values of both $T$ and $U$ 
that gives rise to the traditional flavor symmetry (which leaves the moduli invariant). This leads 
to the traditional flavor group $(D_8\x D_8)/\Z2\cong [32,49]$, see ref.~\cite{Kobayashi:2006wq}. 
The numbers $[32,49]$ correspond to a unique identifier, assigned by the computer program 
GAP~\cite{GAP4}, where the first number (32) gives the order of the group. The finite modular group 
of the $\mathbbm T^2/\Z2$ orbifold is derived in section~\ref{sec:Modular} and turns out to be 
$[144,115]$, the multiplicative closure of mirror symmetry and the $S_3\x S_3$ finite modular 
groups arising from $\SL{2,\Z{}}_T$ and $\SL{2,\Z{}}_U$. If we further include a \CP-like modular 
transformation~\cite{Dent:2001cc,Baur:2019kwi,Novichkov:2019sqv,Baur:2019iai}, this group is 
enhanced to $[288,880]$. Thus, by combining the traditional flavor group and the finite modular 
group we are led to an eclectic flavor group with maximally 2304 elements (without \CP) and 4608 
elements (including \CP).

Typically, only a subgroup of the eclectic flavor group is linearly realized and its size depends 
on the values of the moduli. This leads to an enhancement of the traditional flavor group 
$(D_8\x D_8)/\Z2$ at specific points and hypersurfaces in moduli space: the so-called mechanism of 
{\it local flavor unification}, which is discussed in section~\ref{sec:Localflavorunification}. 
There are two specific configurations of the $\mathbbm T^2/\Z2$ orbifold that deserve special 
attention: the raviolo (at $\vev U=\I$, see figure~\ref{fig:Z2Orbifold}) and the tetrahedron (at 
$\vev U=\exp(\nicefrac{\pi\I}{3})$, see figure~\ref{fig:Z2Tetrahedron}). There, we observe a 
further enhancement of the unified flavor symmetry. The largest linearly realized group is found at 
$\vev T=\vev U=\exp(\nicefrac{\pi\I}{3})$ and turns out to be $[1152,157463]$, which includes 
mirror symmetry and \CP. The landscape of unified flavor symmetries (with \CP) is illustrated in 
figure~\ref{fig:LocalFlavorGroups}. Even for an orbifold as simple as $\mathbbm T^2/\Z2$ we find 
amazingly large flavor groups. In section~\ref{sec:conclusions} we shall summarize our results and 
give an outlook on future research tasks. These should include a full implementation of the 
automorphy factors (in the spirit of refs.~\cite{Nilles:2020tdp,Nilles:2020pr}) and a road-map 
towards an embedding in the six-dimensional case. Some technical results are relegated to three 
appendices.

%%%%%%%%%%%%%%%%%%%%%%%%%%%%%%%%%%%%%%%%%%%%%%%%%%%%%%%%%%%%%%%%%%%%%%%%%%%%%%%%%%%%%%%%%%%%%%%%%%%%%%%%%%%%%%%%%%%%%%%%%%%%%%%%%%%%%%%%%%%%%%%%
\section[Flavor from outer automorphisms of the Z2 space group]{\boldmath Flavor from outer automorphisms of the $\Z{2}$ space group\unboldmath}
\label{sec:Traditional}

\begin{figure*}[t!]
\centering{\includegraphics[width=0.84\linewidth]{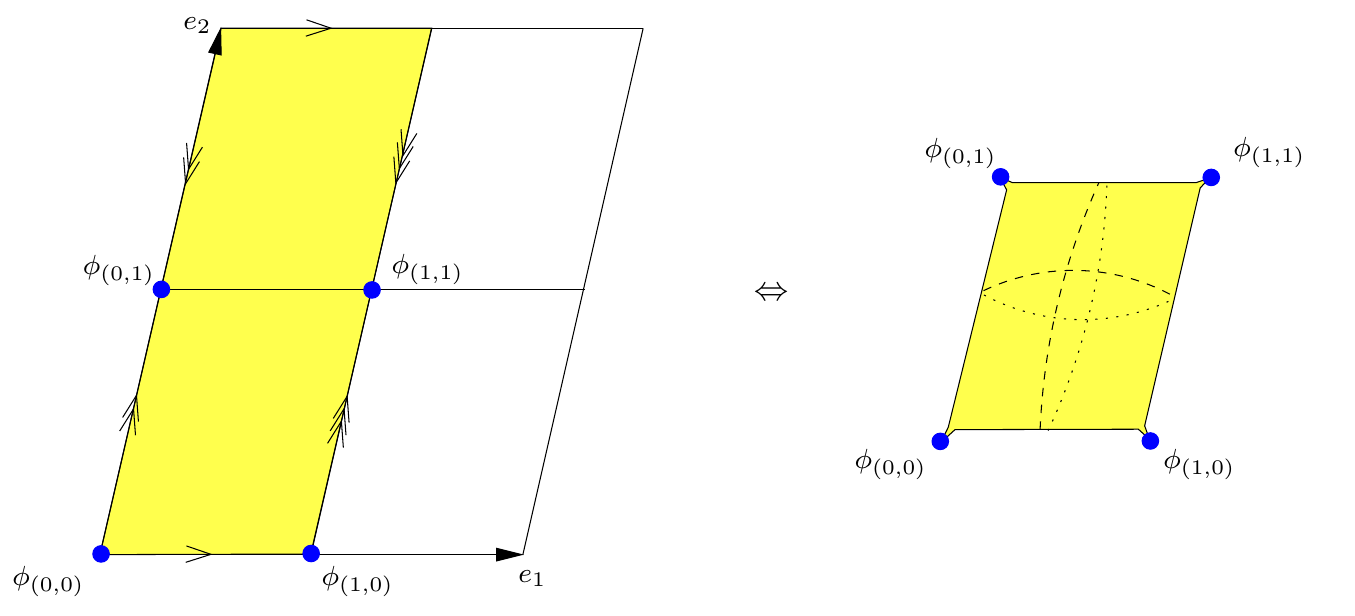}}
\caption{The $\mathbbm{T}^2/\Z{2}$ orbifold. Left: the two-torus $\mathbbm{T}^2$ is defined by a 
two-dimensional lattice, spanned by the vectors $e_1$ and $e_2$. The $\Z{2}$ orbifold twist 
$\theta=-\Id_2$ maps the fundamental domain of $\mathbbm T^2$ to the fundamental domain of the 
orbifold, depicted in yellow. The orbifold action $y \mapsto \theta^k\,y + e\,n$ has four 
inequivalent fixed points, indicated by blue bullets. Localized at these fixed points, there are 
four (left-chiral) twisted strings $(\phi_{(0,0)}, \phi_{(1,0)}, \phi_{(0,1)}, \phi_{(1,1)})$. The 
boundaries of the yellow fundamental domain are identified according to the arrows. Right: after 
identifying the boundaries of the fundamental domain, the $\mathbbm{T}^2/\Z{2}$ orbifold has a 
pillow-like shape with the four fixed points at the corners of the pillow (or raviolo).}
\label{fig:Z2Orbifold}
\end{figure*}

In order to specify the two-dimensional $\mathbbm{T}^2/\Z{2}$ orbifold, we first define the 
geometrical space group $S$. The space group consists of elements $g=(\theta^k, e\,n)\in S$, with 
$k\in\{0,1\}$, that act on the coordinates $y\in\mathbbm{R}^2$ of the extra dimensions as
\begin{equation}\label{eq:actionofspacegroupelement}
y ~\stackrel{g}{\mapsto}~ g\,y ~:=~ \theta^k\,y + e\,n\;,
\end{equation}
where the $\Z{2}$ twist $\theta$ is given by $\theta=-\Id_2$ and $n=(n_1,n_2)^\mathrm{T}\in\Z{}^2$ 
are called winding numbers. One can easily see from eq.~\eqref{eq:actionofspacegroupelement} that 
two space group elements multiply as
\begin{equation}
\label{eq:SpaceGroupGroupLaw}
\big(\theta^k, e\,n\big)\, \big(\theta^{k'}, e\,n'\big) ~=~ \big(\theta^{k+k'}, e\,n + \theta^k e\,n'\big)\;.
\end{equation}
The $2 \times 2$ matrix $e$ (called the geometrical vielbein) consists of two columns $e_1$ and 
$e_2$. These vectors have to be linearly independent, so that they span a two-dimensional lattice 
that defines a two-torus $\mathbbm{T}^2$. Then, the $\mathbbm{T}^2/\Z{2}$ orbifold $\mathbbm{O}$ is 
defined as a quotient space
\begin{equation}
\mathbbm{O} ~:=~ \frac{\mathbbm{R}^2}{S}\,, \qquad\mathrm{where}\quad y ~\sim~ y' \quad\mathrm{if\ there\ exists\ }\quad g~\in~S \quad\mathrm{such\ that}\quad y' ~=~ g\,y\;,
\end{equation}
i.e.\ points $y, y'\in\mathbbm{R}^2$ in extra dimensions are identified if they are related by the 
orbifold action with some space group element $g\in S$. This yields a reduced fundamental 
domain of the orbifold, see figure~\ref{fig:Z2Orbifold}.

Closed strings on $\mathbbm{O}$ are defined by boundary conditions for the string world-sheet 
degrees of freedom~\cite{Dixon:1985jw,Dixon:1986jc,Ibanez:1986tp}, with world-sheet coordinates 
$\tau$ and $\sigma$. Concentrating on the world-sheet bosons $y(\tau,\sigma)$ that describe two 
extra spatial dimensions $y$, a boundary condition for a closed string is given by
\begin{equation}\label{eq:bc}
y(\tau,\sigma+1) ~=~ g\,y(\tau,\sigma) ~=~ \theta^ky(\tau,\sigma) + e\,n\;,
\end{equation}
where $g=(\theta^k, e\,n)\in S$ with $k\in\{0,1\}$ is the so-called constructing element of the 
string. In fact, inequivalent strings correspond not only to constructing elements $g\in S$ but to 
their conjugacy classes $[g] := \{f^{-1}\,g\,f~|~ f\in S\}$, since $y(\tau,\sigma)$ and 
$f\,y(\tau,\sigma)$ are identified on the orbifold for all $f\in S$. If $k=0$ in eq.~\eqref{eq:bc}, 
the string is called untwisted and lives in the bulk of the orbifold. In this case, it can still 
wind around the two-torus depending on its winding numbers $n\in\Z{}^2$. If $k=1$, the string is 
called a twisted string. Then, its center of mass is given by the fixed point $y_g$ of $g\in S$. In 
more detail, $y_g$ denotes the solution of the fixed point equation $g\,y_g=\theta\,y_g+e\,n=y_g$. 
For $g=(\theta,e\,n) \in S$, it reads $y_g = \nicefrac12\,e\,n$. Furthermore, the internal momentum 
of a twisted string vanishes such that a twisted string stays localized at its fixed point $y_g$.

For the $\mathbbm{T}^2/\Z{2}$ orbifold, there are four conjugacy classes of twisted strings with 
constructing elements $g=(\theta, e\, n)=(\theta, n_1\, e_1 + n_2\, e_2)\in S$. They are given by
\begin{equation}
[\big(\theta, n_1\, e_1 + n_2\, e_2\big)] ~=~ \big\{\big(\theta, (n_1+2\tilde{n}_1)\, e_1 + (n_2+2\tilde{n}_2)\, e_2\big)~\in~S ~\big|~ \tilde{n}_1,\tilde{n}_2~\in~\Z{}\big\}\;.
\end{equation}
Hence, winding numbers $n_1$ and $n_2$ of twisted strings are only defined modulo two and we can 
choose $n_1, n_2 \in \{0,1\}$. We denote the four twisted matter fields associated with the four 
conjugacy classes of twisted strings by $\phi_{(n_1, n_2)}$, i.e.
\begin{equation}
\label{eq:conjugacyclasses}
\phi_{(n_1, n_2)} \qquad\Leftrightarrow\qquad [\big(\theta, n_1\, e_1 + n_2\, e_2\big)] \qquad\mathrm{for}\qquad n_1, n_2 ~\in~ \{0,1\}\;.
\end{equation}
Moreover, the matter field $\phi_{(n_1, n_2)}$ is localized at the fixed point 
$y_g = \nicefrac12 (n_1 e_1 + n_2 e_2)$ in $\mathbbm{O}$, as illustrated in 
figure~\ref{fig:Z2Orbifold}.

Discrete flavor symmetries of the effective four-dimensional field theory from strings on orbifolds 
find their origin in the outer automorphisms of the so-called Narain space 
group~\cite{Baur:2019iai,Baur:2019kwi}. Since the Narain construction of strings on orbifolds is 
rather technical, we refer here only to a short discussion in appendix~\ref{app:NarainLattice} and 
to the literature~\cite{Narain:1985jj,Narain:1986am,GrootNibbelink:2017usl}. Still, one can gain 
some insights by considering the outer automorphisms of the geometrical space group $S$ (instead of 
the Narain space group). An outer automorphism of $S$ is given by a transformation 
$h = (\sigma, e\,t)\not\in S$, such that
\begin{equation}\label{eq:OuterOfS}
g ~\stackrel{h}{\longmapsto}~ h^{-1}\,g\,h ~\stackrel{!}{\in}~ S \qquad\mathrm{for\ all\ } g ~\in~ S\;,
\end{equation}
see e.g.\ ref.~\cite{Baur:2019iai} for a similar discussion in the case of a $\mathbbm{T}^2/\Z{3}$ 
orbifold.

\begin{figure}[t!]
\centering
\begin{minipage}{0.4\textwidth}%
\subfloat[h1][Transformation $h_1=\big(\Id_2, \nicefrac{1}{2}\,e_1\big)$]{
\centering{\includegraphics[width=\linewidth]{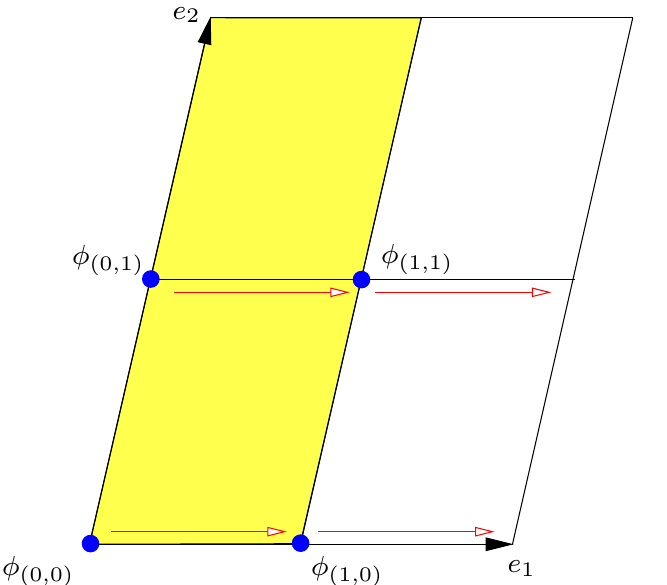}}
\label{fig:LocalFlavorGroupsA}}
\end{minipage}%
\begin{minipage}{0.4\textwidth}%
\subfloat[h2][Transformation $h_2=\big(\Id_2, \nicefrac{1}{2}\,e_2\big)$]{
\centering{\includegraphics[width=\linewidth]{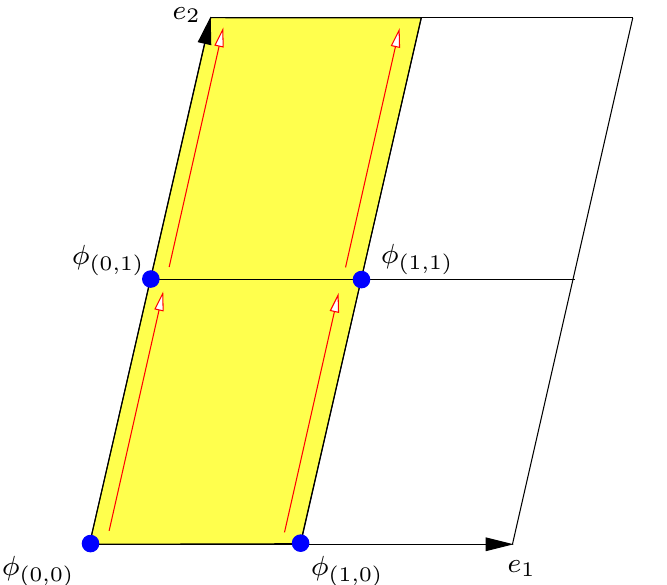}}
\label{fig:LocalFlavorGroupsB}}
\end{minipage}%
\caption{Actions of the outer automorphisms $h_1$ and $h_2$ of the $\mathbbm{T}^2/\Z{2}$ orbifold 
space group on the four twisted matter fields $(\phi_{(0,0)}, \phi_{(1,0)}, \phi_{(0,1)}, \phi_{(1,1)})$.
(a) Under the geometrical translation $h_1 := \big(\Id_2, \nicefrac{1}{2}\,e_1\big)$, matter fields 
get interchanged as $\phi_{(0,n_2)} \leftrightarrow \phi_{(1,n_2)}$ for $n_2\in\{0,1\}$. (b) The 
geometrical translation $h_2 := \big(\Id_2, \nicefrac{1}{2}\,e_2\big)$ interchanges matter fields 
as $\phi_{(n_1,0)} \leftrightarrow \phi_{(n_1,1)}$ for $n_1\in\{0,1\}$.}
\label{fig:SpaceGroupAutomorphims}
\end{figure}

For the $\mathbbm{T}^2/\Z{2}$ orbifold, the outer automorphisms of the geometrical $\Z{2}$ space 
group that leave the moduli unaltered are generated by two translations,
\begin{equation}
h_1 ~:=~ \big(\Id_2, \nicefrac{1}{2}\,e_1\big) \qquad\mathrm{and}\qquad h_2 ~:=~ \big(\Id_2, \nicefrac{1}{2}\,e_2\big)\;.
\end{equation}
In the absence of nontrivial discrete Wilson lines~\cite{Ibanez:1986tp}, they give rise to 
symmetries of the full string construction, see appendix~\ref{app:NarainLattice} for the 
corresponding Narain construction. Then, one is interested in the action $\rho_{\rep{4}}(h)$ of a 
(geometrical) transformation $h\not\in S$ on the four twisted matter fields
\begin{equation}\label{eq:ActionOnPhi}
\begin{pmatrix}\phi_{(0,0)}\\\phi_{(1,0)}\\\phi_{(0,1)}\\\phi_{(1,1)}\end{pmatrix} ~\stackrel{h}{\longmapsto}~ \rho_{\rep{4}}(h)\,\begin{pmatrix}\phi_{(0,0)}\\\phi_{(1,0)}\\\phi_{(0,1)}\\\phi_{(1,1)}\end{pmatrix}\;.
\end{equation}
On the level of constructing elements, one already realizes that, for example, the translation 
$h_1$ acts as
\begin{subequations}
\begin{eqnarray}
\big(\theta, n_1\, e_1 + n_2\, e_2\big) & \stackrel{h_1}{\longmapsto} & \big(\Id_2,\nicefrac{-1}{2}\,e_1\big)\,\big(\theta, n_1\, e_1 + n_2\, e_2\big)\,\big(\Id_2, \nicefrac{1}{2}\,e_1\big)\\ 
&=& \big(\theta, (n_1-1)\, e_1 + n_2\, e_2\big)\;.
\end{eqnarray}
\end{subequations}
Hence, the transformation $h_1$ interchanges $\phi_{(0, n_2)}$ and $\phi_{(1, n_2)}$ for 
$n_2\in\{0,1\}$, see figure~\ref{fig:SpaceGroupAutomorphims}. Similarly, one can show that $h_2$ 
exchanges the twisted matter fields $\phi_{(n_1, 0)}$ and $\phi_{(n_1, 1)}$ for $n_1\in\{0,1\}$. 
This geometrical intuition can be confirmed by a direct computation on twisted string states 
(see appendix~\ref{app:NarainLattice}). We thus find
\begin{equation}\label{eq:OuterOfTwistedStrings}
\rho_{\rep{4}}(h_1) ~=~ \begin{pmatrix}0&1&0&0\\1&0&0&0\\0&0&0&1\\0&0&1&0\end{pmatrix} \quad\mathrm{and}\quad \rho_{\rep{4}}(h_2) ~=~ \begin{pmatrix}0&0&1&0\\0&0&0&1\\1&0&0&0\\0&1&0&0\end{pmatrix}\;.
\end{equation}
These transformations generate a $\Z{2}\times\Z{2}$ Abelian flavor symmetry.

In addition, there are string selection rules restricting the ability of strings on orbifolds to 
join and split~\cite{Hamidi:1986vh}. The associated symmetry can be determined in two ways: i) 
as the Abelianization of the space group $S$~\cite{Ramos-Sanchez:2018edc}, or ii) as additional 
outer automorphisms of the Narain space group. Combined with the geometrical transformations $h_1$ 
and $h_2$, which exchange orbifold fixed points pairwise, these string selection rules yield a 
non-Abelian flavor symmetry as follows: in the $\mathbbm T^2/\Z2$ orbifold, the string selection 
rules give rise to a $\Z{2}\times\Z{2}\times\Z{2}$ symmetry, under which twisted matter fields 
transform as
\begin{subequations}\label{eq:SelectionRulesOfTwistedStrings}
\begin{eqnarray}
\phi_{(n_1,n_2)} &\stackrel{h_3}{\longmapsto}& (-1)^{n_1}\,\phi_{(n_1,n_2)}\qquad\Rightarrow\qquad \rho_{\rep{4}}(h_3) ~=~ \begin{pmatrix} 1&0&0&0\\0&-1&0&0\\0&0& 1&0\\0&0&0&-1\end{pmatrix}\;,\label{eq:SGn1}\\
\phi_{(n_1,n_2)} &\stackrel{h_4}{\longmapsto}& (-1)^{n_2}\,\phi_{(n_1,n_2)}\qquad\Rightarrow\qquad \rho_{\rep{4}}(h_4) ~=~ \begin{pmatrix} 1&0&0&0\\0& 1&0&0\\0&0&-1&0\\0&0&0&-1\end{pmatrix}\;,\label{eq:SGn2}\\
\phi_{(n_1,n_2)} &\stackrel{h_5}{\longmapsto}& -\phi_{(n_1,n_2)} \,\,\qquad\qquad\Rightarrow\qquad \rho_{\rep{4}}(h_5) ~=~ \begin{pmatrix}-1&0&0&0\\0&-1&0&0\\0&0&-1&0\\0&0&0&-1\end{pmatrix}\;.\label{eq:PG}
\end{eqnarray}
\end{subequations}
Here, $h_3$ and $h_4$ are connected to the $\Z{2}\times\Z{2}$ space group selection rule for 
twisted strings, while $h_5$ is associated with the \Z{2} point group selection rule. Then, as 
first shown in ref.~\cite{Kobayashi:2006wq}, the full traditional flavor group (without 
$R$-symmetry) of the $\mathbbm{T}^2/\Z{2}$ orbifold, at a generic point in moduli space, is 
generated by the transformations~\eqref{eq:OuterOfTwistedStrings} 
and~\eqref{eq:SelectionRulesOfTwistedStrings}, resulting in
\begin{equation}\label{eq:TFS}
\frac{\left(D_8 \times D_8\right)}{\Z{2}} ~\cong~ [32,49]\;.
\end{equation}
Here, $D_8 \cong [8,3]$ denotes the dihedral group of order 8 (sometimes also called $D_4$ 
using a different naming convention). Furthermore, the first $D_8$ factor in eq.~\eqref{eq:TFS} is 
generated by $h_1$ and $h_3$. This $D_8$ is associated with the $e_1$ direction of the orbifold, 
cf.\ ref.~\cite{Kobayashi:2004ya}. The second $D_8$ in eq.~\eqref{eq:TFS} is generated by $h_2$ and 
$h_4$ and is associated with $e_2$. Note that the transformation $h_5$ in eq.~\eqref{eq:PG}, linked 
to the point group selection rule, is not an independent generator of the traditional flavor group 
eq.~\eqref{eq:TFS} as it can be written as
\begin{equation}
\rho_{\rep{4}}(h_5) ~=~ \big(\rho_{\rep{4}}(h_1)\, \rho_{\rep{4}}(h_3)\big)^2 ~=~ \big(\rho_{\rep{4}}(h_2)\, \rho_{\rep{4}}(h_4)\big)^2\;.
\end{equation}
This identity gives rise to the $\Z{2}$ quotient in eq.~\eqref{eq:TFS}. Moreover, the four twisted 
matter fields $(\phi_{(0,0)}, \phi_{(1,0)}, \phi_{(0,1)}, \phi_{(1,1)})^\mathrm{T}$ build a 
four-dimensional, irreducible, and faithful representation of the traditional flavor group eq.~\eqref{eq:TFS}. 

Even though we have discussed the origin of the traditional flavor symmetry $(D_8\times D_8)/\Z{2}$ 
based on the outer automorphisms of the geometrical $\Z{2}$ space group, the results can be 
confirmed using the full Narain approach, see appendices~\ref{app:NarainLattice} and~\ref{app:Lutowski}. 
Even more, the full Narain space group and its outer automorphisms reveal a common origin of all discrete 
symmetries for strings on orbifolds, giving rise to the eclectic flavor symmetry that consists of 
traditional flavor, modular, $\CP$ and $R$-symmetries. In the next section, we will analyze in 
detail the modular symmetries and their finite modular groups that arise in the $\mathbbm T^2/\Z2$ 
orbifold.

%%%%%%%%%%%%%%%%%%%%%%%%%%%%%%%%%%%%%%%%%%%%%%%%%%%%%%%%%%%%%%%%%%%%%%%%%%%%%%%%%%%%%%%%%%%%%%%%%%%%%%%%%%%%%%%%%%%%%%%%%%%%%%%%%%%%%%%%%
\section{Flavor from modular symmetries}
\label{sec:Modular}

In general, the deformations of a two-dimensional torus used to compactify a string theory can be 
parameterized by a complex structure modulus $U$ and a K\"ahler modulus $T$. The complex structure 
modulus $U$ can be interpreted geometrically as the shape of the torus, while the K\"ahler modulus 
$T$ gives the overall size of the torus and the value of the anti-symmetric $B$-field 
background. Moreover, a toroidal compactification exhibits several symmetries that act nontrivially 
on these moduli (see e.g.\ ref.~\cite{Baur:2019iai}):
\begin{subequations}\label{eq:SLZ2UandT}
\begin{eqnarray}
U & \stackrel{\hat{C}_\mathrm{S}}{\longmapsto} & -\frac{1}{U} \hspace{1.05cm}\mathrm{and}\qquad T ~\stackrel{\hat{C}_\mathrm{S}}{\longmapsto}~ T\;,\label{eq:CS}\\
U & \stackrel{\hat{C}_\mathrm{T}}{\longmapsto} & U+1                \qquad\mathrm{and}\qquad T ~\stackrel{\hat{C}_\mathrm{T}}{\longmapsto}~ T\;,\label{eq:CT}\\
U & \stackrel{\hat{K}_\mathrm{S}}{\longmapsto} & U       \,\;\;\quad\qquad\mathrm{and}\qquad T ~\stackrel{\hat{K}_\mathrm{S}}{\longmapsto}~ -\frac{1}{T}\;,\\
U & \stackrel{\hat{K}_\mathrm{T}}{\longmapsto} & U       \,\;\;\quad\qquad\mathrm{and}\qquad T ~\stackrel{\hat{K}_\mathrm{T}}{\longmapsto}~ T+1\;.
\end{eqnarray}
\end{subequations}
These transformations are actually not defined on the level of the moduli but on the level of outer 
automorphisms of the Narain space group, see appendix~\ref{app:Out}. Then, $\hat{C}_\mathrm{S}$ and 
$\hat{C}_\mathrm{T}$ generate a modular group $\SL{2,\Z{}}_U$ associated with the complex structure 
modulus $U$, while $\hat{K}_\mathrm{S}$ and $\hat{K}_\mathrm{T}$ are the generators of another 
factor $\SL{2,\Z{}}_T$ associated with the K\"ahler modulus $T$. Note that the two factors of 
$\SL{2,\Z{}}$ share a common element: $\mathrm{C}:=(\hat{C}_\mathrm{S})^2=(\hat{K}_\mathrm{S})^2$. 
Even though $\mathrm{C}$ acts trivially on both moduli, it can in principle still act nontrivially 
on matter fields, see ref.~\cite{Baur:2019iai,Nilles:2020kgo} (and also ref.~\cite{Ohki:2020bpo}). 
In addition, there are two special transformations
\begin{subequations}\label{eq:MandSigmaStar}
\begin{eqnarray}
U & \stackrel{\hat{M}}{\longmapsto}      & T     \,\;\;\quad\qquad\mathrm{and}\qquad T ~\stackrel{\hat{M}}{\longmapsto}~        U\;,\label{eq:Mirror}\\
U & \stackrel{\hat\Sigma_*}{\longmapsto} & -\bar{U}   \,\;\;\qquad\mathrm{and}\qquad T ~\stackrel{\hat\Sigma_*}{\longmapsto}~ -\bar{T}\;.
\end{eqnarray}
\end{subequations}
The former is the origin of the so-called mirror symmetry that interchanges $\SL{2,\Z{}}_U$ and 
$\SL{2,\Z{}}_T$, while the latter induces a \CP-like transformation, see 
refs.~\cite{Dent:2001cc,Baur:2019kwi} and~\cite{Novichkov:2019sqv}.

In principle, performing an orbifold of a torus can stabilize some moduli geometrically. 
Consequently, some of the symmetry transformations generated by eqs.~\eqref{eq:SLZ2UandT} 
and~\eqref{eq:MandSigmaStar} can be broken by the orbifolding. For example, in the case of a 
$\mathbbm{T}^2/\Z{3}$ orbifold sector the $U$ modulus needs to be stabilized, e.g.\ at 
$\vev U=\exp(\nicefrac{2\pi\I}{3})$, and the unbroken modular symmetry after $\Z{3}$ orbifolding is 
generated by $\hat{K}_\mathrm{S}$, $\hat{K}_\mathrm{T}$, the $R$-symmetry 
$\hat{C}_\mathrm{S}\,\hat{C}_\mathrm{T}$ and the \CP-like transformation 
$\hat{K}_* := \hat{C}_\mathrm{S}\, \hat{C}_\mathrm{T}\, \hat{C}_\mathrm{S}\, \hat{\Sigma}_*$ (i.e.\ 
by those modular transformations that leave $\vev U=\exp(\nicefrac{2\pi\I}{3})$ invariant). In 
contrast, the $\mathbbm{T}^2/\Z{2}$ orbifold with $\theta = -\Id_2$ is equipped with both moduli: 
the complex structure modulus $U$ and the K\"ahler modulus $T$. Each of them remains associated 
with its own unbroken modular group, $\SL{2,\Z{}}_U$ and $\SL{2,\Z{}}_T$ for $U$ and $T$, 
respectively. Moreover, the transformations~\eqref{eq:MandSigmaStar} remain symmetries after the 
torus has been modded out by the $\Z{2}$ orbifold action $\theta = -\Id_2$. Hence, the 
$\mathbbm{T}^2/\Z{2}$ orbifold gives a simple example of a string setup with multiple modular 
symmetries (see e.g.\ ref.~\cite{deMedeirosVarzielas:2019cyj}), with the extension by mirror 
symmetry eq.~\eqref{eq:Mirror} that interchanges both moduli.

As shown in refs.~\cite{Nilles:2020tdp,Nilles:2020pr}, the modular group $\SL{2,\Z{}}_U$ of the 
complex structure modulus acts geometrically on the compact dimensions. In particular, the 
$\SL{2,\Z{}}_U$ generators $\hat{C}_\mathrm{S}$ and $\hat{C}_\mathrm{T}$ act on the $\mathbbm T^2$ 
basis vectors $e_i$ according to
\begin{equation}
\label{eq:SL2ZOnT2Basis}
e_1  \stackrel{\hat{C}_\mathrm{S}}{\longrightarrow} e'_1 = -e_2 \,,\quad e_2 \stackrel{\hat{C}_\mathrm{S}}{\longrightarrow} e'_2 = e_1\,,\quad\mathrm{and}\quad
e_1  \stackrel{\hat{C}_\mathrm{T}}{\longrightarrow} e'_1 =  e_1 \,,\quad e_2 \stackrel{\hat{C}_\mathrm{T}}{\longrightarrow} e'_2 = e_1+e_2\,.
\end{equation}
This can be confirmed easily by e.g. considering each two-dimensional vector $e_i$ as a complex 
number, i.e.\ $e_1, e_2\in\mathbbm{C}$, and setting $U=\nicefrac{e_2}{e_1}$, see 
ref.~\cite{Kikuchi:2020frp}. Then, eq.~\eqref{eq:SL2ZOnT2Basis} reproduces eqs.~\eqref{eq:CS} 
and~\eqref{eq:CT}.

\begin{figure}[t]
\centering
\begin{minipage}{0.4\textwidth}%
\subfloat[S transformation][Modular $\mathrm{S}$ transformation.]{
\centering{\includegraphics[width=\linewidth]{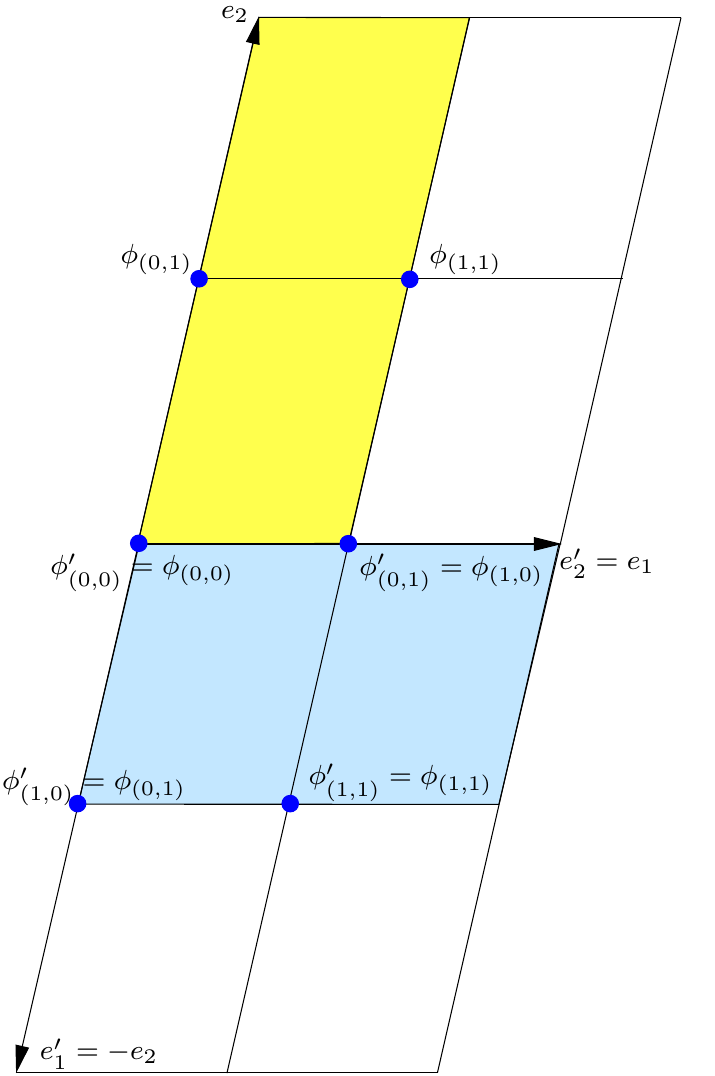}}
\label{fig:SL2ZUS}}
\end{minipage}%
\begin{minipage}{0.608\textwidth}%
\subfloat[T transformation][Modular $\mathrm{T}$ transformation.]{
\centering{\includegraphics[width=\linewidth]{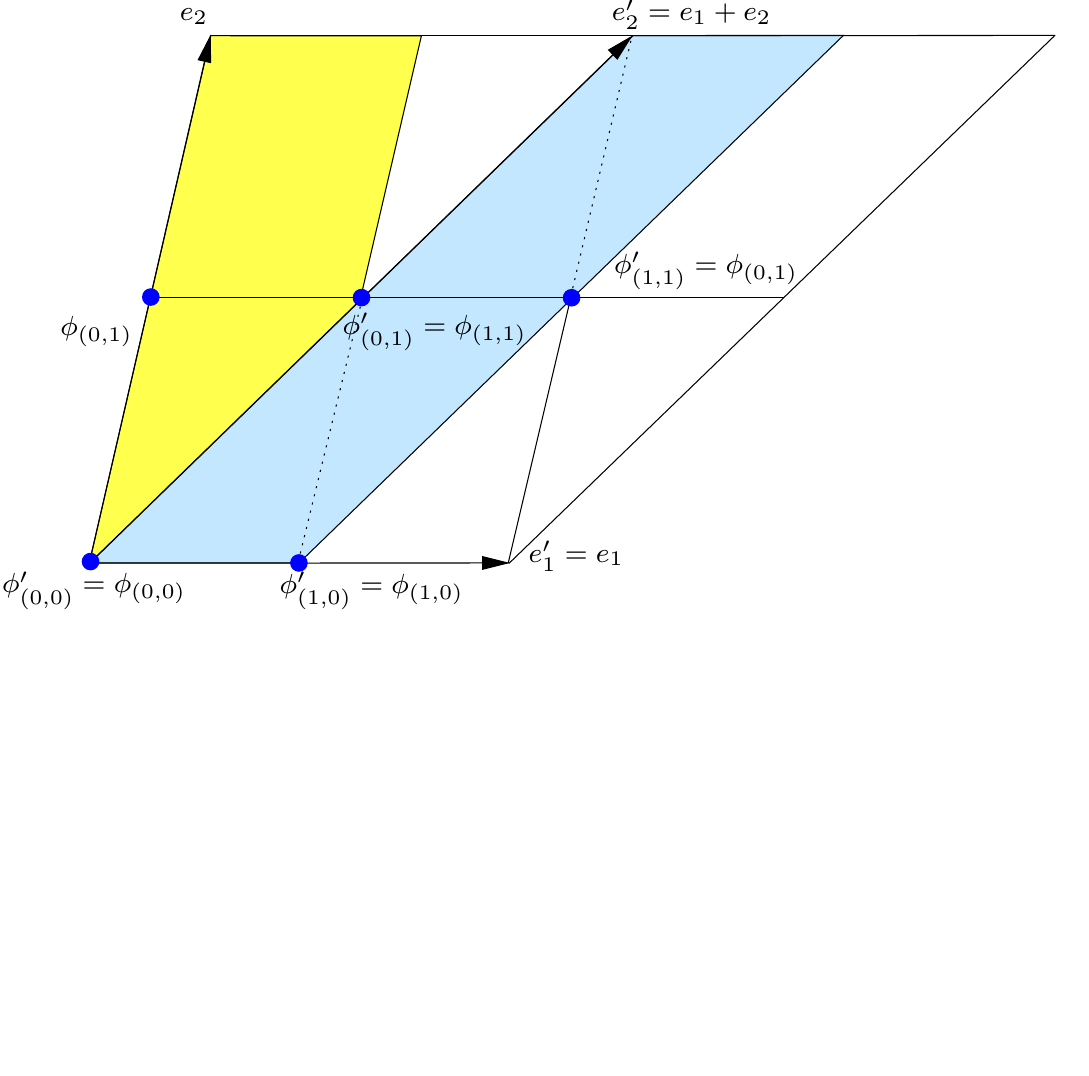}}
\label{fig:SL2ZUT}}
\end{minipage}%
\caption{The action of the $\SL{2,\Z{}}_U$ generators $\hat C_\mathrm{S}$ (a) and 
$\hat C_\mathrm{T}$ (b) associated with the complex structure modulus $U$ on the 
$\mathbbm{T}^2/\Z{2}$ orbifold. From the action of the $\SL{2,\Z{}}_U$ generators on the 
$\mathbbm T^2$ basis, eq.~\eqref{eq:SL2ZOnT2Basis}, the original (yellow) fundamental domain of the 
orbifold is mapped to an equivalent but different (blue) region. Hence, twisted matter 
fields $\phi_{(n_1,n_2)}$ get interchanged according to eqs.~\eqref{eq:defGammaUAction} 
and~\eqref{eq:SL2ZUOfTwistedStrings}.}
\label{fig:SL2ZU}
\end{figure}

Next, we are interested in the action of $\SL{2,\Z{}}_U$ on the four twisted matter fields 
$\phi_{(n_1, n_2)}$, i.e.\ for $\gamma_U\in\SL{2,\Z{}}_U$ we want to identify 
$\rho_{\rep{4}}(\gamma_U)$ defined as
\begin{equation}
\label{eq:defGammaUAction}
\begin{pmatrix}\phi_{(0,0)}\\\phi_{(1,0)}\\\phi_{(0,1)}\\\phi_{(1,1)}\end{pmatrix} ~\stackrel{\gamma_U}{\longmapsto}~ \begin{pmatrix}\phi'_{(0,0)}\\\phi'_{(1,0)}\\\phi'_{(0,1)}\\\phi'_{(1,1)}\end{pmatrix} ~=~ \rho_{\rep{4}}(\gamma_U)\,\begin{pmatrix}\phi_{(0,0)}\\\phi_{(1,0)}\\\phi_{(0,1)}\\\phi_{(1,1)}\end{pmatrix}\;,
\end{equation}
without taking the automorphy factors $(c_U\,U+d_U)^{n_U}$ of $\gamma_U$ with modular weight $n_U$ into 
account. Then, one can use the geometrical $\SL{2,\Z{}}_U$ transformations~\eqref{eq:SL2ZOnT2Basis} 
as illustrated in figure~\ref{fig:SL2ZU} in order to obtain the matrix representations of the 
modular $\mathrm{S}$ and $\mathrm{T}$ transformations, given by $\hat{C}_\mathrm{S}$ and 
$\hat{C}_\mathrm{T}$, for the four twisted matter fields $\phi_{(n_1, n_2)}$. The results read
\begin{equation}
\label{eq:SL2ZUOfTwistedStrings}
\rho_{\rep{4}}(\hat{C}_\mathrm{S}) ~=~ \begin{pmatrix}1&0&0&0\\0&0&1&0\\0&1&0&0\\0&0&0&1\end{pmatrix} \quad\mathrm{and}\quad \rho_{\rep{4}}(\hat{C}_\mathrm{T}) ~=~ \begin{pmatrix}1&0&0&0\\0&1&0&0\\0&0&0&1\\0&0&1&0\end{pmatrix}\;.
\end{equation}
One can check that these representation matrices $\rho_{\rep{4}}(\hat C_\mathrm{S})$ and 
$\rho_{\rep{4}}(\hat C_\mathrm{T})$ generate a so-called finite modular group $S_3\cong[6,1]$: 
even though the symmetry is $\SL{2,\Z{}}_U$, twisted matter fields transform under $\SL{2,\Z{}}_U$ 
in unitary representations of $S_3$. We denote this factor by $S_3^U$ as it is associated with the 
complex structure modulus $U$.

Also the mirror symmetry $\hat{M}$ has a non-trivial action on twisted matter fields. It turns out that 
it can be represented by the matrix
\begin{equation}
\label{eq:repM}
\rho_{\rep{4}}(\hat{M}) ~=~ \frac{1}{\sqrt{2}}\begin{pmatrix}0&0&-1&1\\0&0&1&1\\1&-1&0&0\\-1&-1&0&0\\ \end{pmatrix}\;,
\end{equation}
see appendix~\ref{app:mirror}. Interestingly, one can verify easily that
\begin{equation}
\label{eq:squaredrepM}
\left(\rho_{\rep{4}}(\hat{M})\right)^2 ~=~ -\Id_4\;.
\end{equation}
Hence, $\rho_{\rep{4}}(\hat{M})$ is of order 4. Moreover, eq.~\eqref{eq:squaredrepM} shows that, 
although $\hat M^2$ acts trivially on the moduli, it acts nontrivially on twisted matter fields: in 
fact, $(\rho_{\rep{4}}(\hat{M}))^2$ acts like the traditional flavor transformation 
$\rho_{\rep{4}}(h_5)=-\Id_4$ associated with the point group selection rule, see eq.~\eqref{eq:PG}. 
Consequently, the finite modular group can not be disentangled from the traditional flavor group 
completely: the element $\hat{M}^2$ belongs to both groups. The situation is similar to the 
$\mathbbm T^2/\Z3$ orbifold, where the modular $\mathrm{S}$ transformation squared equals a 
traditional flavor transformation: $(\hat{C}_\mathrm{S})^2=(\hat{K}_\mathrm{S})^2=\mathrm{C}$ from 
$\Delta(54)$, see ref.~\cite{Baur:2019iai}.

Then, we can translate the finite modular group $S_3^U$ of the complex structure modulus $U$ to the 
one of the K\"ahler modulus $T$. Using eq.~\eqref{eq:mirrorsymmetry} from appendix~\ref{app:Out}, we 
obtain
\begin{subequations}
\label{eq:SL2ZTOfTwistedStrings}
\begin{eqnarray}
\rho_{\rep{4}}(\hat{K}_\mathrm{S}) & := & \rho_{\rep{4}}(\hat{M})\,\rho_{\rep{4}}(\hat{C}_\mathrm{S})\,\rho_{\rep{4}}(\hat{M})^{-1} ~=~  \frac{1}{2}\begin{pmatrix}1&1&1&1\\1&1&-1&-1\\1&-1&1&-1\\1&-1&-1&1\\ \end{pmatrix}\;,\\
\rho_{\rep{4}}(\hat{K}_\mathrm{T}) & := & \rho_{\rep{4}}(\hat{M})\,\rho_{\rep{4}}(\hat{C}_\mathrm{T})\,\rho_{\rep{4}}(\hat{M})^{-1} ~=~  \begin{pmatrix}-1&0&0&0\\0&1&0&0\\0&0&1&0\\0&0&0&1\\ \end{pmatrix}\;,
\end{eqnarray}
\end{subequations}
see ref.~\cite{Lauer:1990tm}. We denote the resulting finite modular group associated with the 
K\"ahler modulus $T$ by $S_3^T$.

The final transformation from the list of generators given in eqs.~\eqref{eq:SLZ2UandT} 
and~\eqref{eq:MandSigmaStar} is the \CP-like transformation $\hat\Sigma_*$. It acts on twisted 
matter fields as
\begin{equation}
\label{eq:SigmaOnTwistedStrings}
\begin{pmatrix}\phi_{(0,0)}\\\phi_{(1,0)}\\\phi_{(0,1)}\\\phi_{(1,1)}\end{pmatrix} ~\stackrel{\hat\Sigma_*}{\longmapsto}~ \begin{pmatrix}\bar\phi_{(0,0)}\\\bar\phi_{(1,0)}\\\bar\phi_{(0,1)}\\\bar\phi_{(1,1)}\end{pmatrix}\;,
\end{equation}
where we suppress the spacetime dependencies. Hence, if one considers \CP-like transformations, it 
is beneficial to extend the $4\times 4$ representation matrices to $8\times 8$ matrices acting on 
the eight-dimensional vector $(\Phi,\bar{\Phi})^\mathrm{T}$ of twisted matter fields 
$\Phi:=(\phi_{(0,0)},\phi_{(1,0)},\phi_{(0,1)},\phi_{(1,1)})^\mathrm{T}$ and their \CP-partners.

In summary, the matrices $\rho_{\rep{4}}(\hat{K}_\mathrm{S})$, $\rho_{\rep{4}}(\hat{K}_\mathrm{T})$, 
$\rho_{\rep{4}}(\hat{C}_\mathrm{S})$ and $\rho_{\rep{4}}(\hat{C}_\mathrm{T})$ generate the finite 
modular group
\begin{equation}
S_3^T \x S_3^U
\end{equation}
of order $6 \times 6 = 36$. Combined with the $\Z4$ mirror element $\rho_{\rep{4}}(\hat{M})$, we 
obtain the finite modular group without $\CP$
\begin{equation}
\label{eq:GenericFiniteModularGroupNoCP}
\left(S_3^T \x S_3^U\right)\rtimes\Z4^{\hat M} ~\cong~ [144,115]\;,
\end{equation}
which is of order $36 \times 4 = 144$. We observe, as a side remark, that this finite modular group is
related to the group of outer automorphisms of the traditional flavor group,
\begin{equation}
\mathrm{Out}\big((D_8\x D_8)/\Z2\big) ~\cong~ [72,40] ~\cong~ [144,115]/\Z{2}\;,
\end{equation}
where the \Z2 on the right-hand side is generated by $\hat M^2$.
Moreover, by including $\CP$ we get 
$[288, 880] \cong [144,115] \times \Z{2}$, which is the maximal finite modular group of the 
$\mathbbm T^2/\Z2$ orbifold.

Next, we combine the finite modular group with the traditional flavor group $(D_8\x D_8)/\Z2$ and 
construct the eclectic flavor group $G_\text{eclectic}$. The traditional flavor 
group is a normal subgroup of $G_\text{eclectic}$, as expected from 
the general framework of eclectic flavor groups~\cite{Nilles:2020nnc}.
However, since $(D_8\x D_8)/\Z2$ and the finite modular group share 
the common element $(\rho_{\rep{4}}(\hat{M}))^2 = \rho_{\rep{4}}(h_5)$, the eclectic flavor group 
is not a semi-direct product of these two factors. In the case without \CP, the finite modular group 
is $(S_3^T \x S_3^U)\rtimes\Z4^{\hat M}$ and $G_\text{eclectic}$ turns out to be of order 2304. This 
order can be understood easily since $(144 \times 32)/2 = 2304$ using the fact that 
$(\rho_{\rep{4}}(\hat{M}))^2$ belongs to both factors. As a side remark, note that all finite groups of 
order 2304 have been classified in ref.~\cite{Eick_2014}. In appendix~\ref{app:irreps} we examine 
the representation $\rho_{\rep{4}}$ of twisted matter fields with respect to $G_\text{eclectic}$ 
and its various subgroups. In addition, if we include $\CP$, the eclectic flavor group gets 
enhanced further to a group of order 4608.

%%%%%%%%%%%%%%%%%%%%%%%%%%%%%%%%%%%%%%%%%%%%%%%%%%%%%%%%%%%%%%%%%%%%%%%%%%%%%%%%%%%%%%%%%%%%%%%%%%%%%%%%%%%%%%%%%%%%%%%%%%%%%%%%%%%%%%%%%%
\section{Local flavor unification}
\label{sec:Localflavorunification} 

As discussed in section~\ref{sec:Traditional}, for generic values of the moduli, the traditional 
flavor symmetry of the $\mathbbm T^2/\Z2$ orbifold is $(D_8\x D_8)/\Z2\cong[32,49]$, cf.\ 
eq.~\eqref{eq:TFS}, associated with the discrete symmetries of the theory that do not affect the 
moduli. 

On the other hand, as explained in section~\ref{sec:Modular}, omitting the \CP-like transformation 
$\hat\Sigma_*$ in a first step, the finite modular group of the $\mathbbm T^2/\Z2$ orbifold is 
$(S_3^T\x S_3^U)\rtimes\Z4^{\hat M}\cong [144,115]$. This group can be constructed by the order 4 
generator $\rho_{\rep 4}(\hat M)$ associated with the mirror transformation $\hat{M}$, see 
eq.~\eqref{eq:repM}, and the $S_3^T\x S_3^U$ finite modular transformations arising from 
$\SL{2,\Z{}}_T$ and $\SL{2,\Z{}}_U$. These modular symmetries are, in general, independent of the 
traditional flavor symmetry because traditional flavor transformations act trivially on the moduli 
$T$ and $U$. In contrast, modular transformations act by definition nontrivially on these moduli. 
We assume now that the moduli are fixed at some vacuum expectation values (vevs) $(\vev{T},\vev{U})$, 
see e.g.\ ref.~\cite{Abe:2020vmv}. Then, we have to distinguish between two cases: first, there are 
modular transformations that do not leave the moduli vevs invariant. They are broken spontaneously. 
Second, some modular transformations may leave the moduli vevs invariant. They build the so-called 
{\it stabilizer subgroup}
\begin{equation}
\label{eq:stabilizerDefinition}
H_{(\vev{T},\vev{U})} ~:=~ \big\langle ~\gamma ~\big|~ \gamma ~\in~ \Xi \quad
     \text{with}\quad \gamma(\vev{T}) ~=~ \vev{T} \;\;\mathrm{and}\;\; \gamma(\vev{U}) ~=~ \vev{U}~\big\rangle\;,
\end{equation}
which depends on the moduli vevs $\vev{T}$ and $\vev{U}$. By definition, $H_{(\vev{T},\vev{U})}$ 
is a subgroup of $\Xi=\mathrm{O}_{\hat\eta}(2,2,\Z{})/\Z{2}$, see eq.~\eqref{eq:LutowskiStabilizer}. 
As detailed in appendix~\ref{app:Lutowski}, $\Xi$ is given by the full modular group 
$\mathrm{O}_{\hat\eta}(2,2,\Z{})$ defined in eq.~\eqref{eq:OEtaHat22Z} divided by the \Z2 point 
group, where $\mathrm{O}_{\hat\eta}(2,2,\Z{})$ comprises $\SL{2,\Z{}}_T$, $\SL{2,\Z{}}_U$, the 
mirror transformation $\hat M$, and the \CP-like transformation $\hat\Sigma_*$. If the stabilizer 
subgroup at a special point in moduli space is nontrivial, i.e.\ $H_{(\vev{T},\vev{U})}\neq\{\Id\}$, 
then its elements remain unbroken by the vevs. Due to their trivial action on the moduli vevs and 
nontrivial action on matter fields, these unbroken transformations enhance the traditional flavor 
group to a so-called {\it unified flavor group}
\begin{equation}
\label{eq:enhancement}
   \frac{D_8\x D_8}{\Z2} ~\cup~ H_{(\vev{T},\vev{U})}\;,
\end{equation}
which results from the multiplicative closure of the universal traditional flavor group and
the stabilizer subgroup $H_{(\vev{T},\vev{U})}$. 

Thus, given the representation of the elements of the traditional flavor group in the field basis, 
determining the unified flavor group at the point $(\vev{T},\vev{U})$ requires to know the matrix 
representation $\rho(\gamma)$ associated with the action of the element $\gamma$ of the stabilizer 
subgroup $H_{(\vev{T},\vev{U})}$ on the orbifold matter fields $\Phi$. For the four twisted matter 
fields of the $\mathbbm T^2/\Z2$ orbifold, $\rho(\gamma)$ can be built from the $\rho_{\rep 4}$ 
representation matrices given in eqs.~\eqref{eq:SL2ZUOfTwistedStrings},~\eqref{eq:repM} 
and~\eqref{eq:SL2ZTOfTwistedStrings}.

However, since string matter fields carry (fractional) modular weights $(n_T,n_U)$ of 
$\SL{2,\Z{}}_T$ and $\SL{2,\Z{}}_U$, the matrix representation $\rho(\gamma)$ of a modular 
transformation $\gamma$ at the point $(\vev{T},\vev{U})$ in moduli space is accompanied by 
so-called {\it automorphy factors} of the form $(c_T\vev T +d_T)^{n_T}(c_U\vev U +d_U)^{n_U}$, 
where $c_T,d_T,c_U,d_U$ are integers parametrizing the transformation 
$\gamma\in H_{(\vev{T},\vev{U})}$. As discussed in ref.~\cite{Nilles:2020tdp}, these automorphy 
factors evaluated at the vevs $(\vev{T},\vev{U})$ are discrete phases. This fact can i) change the 
order of the flavor symmetry associated with $\gamma$, and/or ii) reveal that $\gamma$ acts as a 
discrete $R$-symmetry, which in general promotes the unified flavor group at $(\vev{T},\vev{U})$ to 
a non-Abelian discrete $R$-symmetry of $\mathcal{N}=1$ supersymmetry~\cite{Chen:2013dpa}. However, 
for the sake of clarity and simplicity, we ignore the automorphy factors in the following. Hence, 
hereafter we shall provide the actual groups of the unified flavor symmetries only in cases where 
the automorphy factors do not affect the results, but shall give in all cases the generators of the 
corresponding stabilizer subgroups. The subtleties about the consequences of the automorphy factors 
as well as their relevance in six-dimensional orbifolds shall be discussed in detail later in 
ref.~\cite{Baur:2020pr}.

A first example of a unified flavor symmetry arises at $\vev{T}=\vev{U}$ with generic vev. This 
is a two-dimensional hypersurface in four-dimensional (real) moduli space. At these points, the 
mirror symmetry transformation $\hat{M}$, acting on the moduli as 
$T\stackrel{\hat{M}}{\longleftrightarrow}U$, generates a \Z2 stabilizer subgroup. Considering the 
representation $\rho_{\rep{4}}(\hat{M})$ of $\hat M$ for twisted matter fields, eq.~\eqref{eq:repM}, 
one finds that the traditional flavor group 
\begin{equation}
[32,49]\cong(D_8\x D_8)/\Z2 \qquad\text{enhances to}\qquad [64,257] \quad\text{at}\quad \vev{T}~=~\vev{U}\;.
\end{equation}
Note that, although the order of $\rho_{\rep{4}}(\hat{M})$ is 4, the order of the flavor group is 
enhanced only by a factor of 2 because $(\rho_{\rep{4}}(\hat{M}))^2=-\Id_4$ is also included in the 
traditional flavor group, cf.~eq.~\eqref{eq:squaredrepM}.

Mirror symmetry $\hat M$ helps to simplify the study of unified flavor symmetries at special points 
in moduli space. We know that the mirror transformation $\hat M$ maps 
\begin{equation}
U ~\stackrel{\hat M}{\longleftrightarrow} ~T \quad,\quad \hat{K}_\mathrm{S} ~\stackrel{\hat M}{\longleftrightarrow}~ \hat{C}_\mathrm{S} \quad\mathrm{and}\quad \hat{K}_\mathrm{T} ~\stackrel{\hat M}{\longleftrightarrow}~ \hat{C}_\mathrm{T}\;,
\end{equation}
cf.~eq.~\eqref{eq:mirrorsymmetry} in appendix~\ref{app:Out}. Since $\hat{M}$ is a symmetry of the theory, 
there is an equivalence between the unbroken modular symmetry at the point $(T,U)=(\vev T,\vev U)$ 
in moduli space and its mirror dual at the point $(T,U)=(\vev U,\vev T)$. This implies that the 
stabilizer subgroups satisfy the isomorphism
\begin{equation}
\label{eq:stabilizerIsomorphism}
H_{(T=\vev T,\,U=\vev U)}~\cong~ H_{(T=\vev U,\,U=\vev T)}\;,
\end{equation}
using $\hat M\, \gamma\, \hat M^{-1} \in H_{(T=\vev U,\,U=\vev T)}$ for all 
$\gamma\in H_{(T=\vev T,\,U=\vev U)}$ and $\hat M\, h\, \hat M^{-1} \in (D_8\times D_8)/\Z{2}$ for 
all $h\in (D_8\times D_8)/\Z{2}$. Consequently, the associated unified flavor symmetries, resulting 
from combining these stabilizer subgroups with the universal traditional flavor group, as 
prescribed by eq.~\eqref{eq:enhancement}, are isomorphic too. This implies that mirror symmetry 
$\hat M$ halves the fundamental domain in the full moduli space and we must not explore both 
isomorphic cases eq.~\eqref{eq:stabilizerIsomorphism} independently. Thus, we shall explore only 
those symmetry enhanced points and hypersurfaces in moduli space that are associated with a 
geometric interpretation of $U$. Then, for each case of $\langle U \rangle$, we identify the symmetry 
enhanced points in the $T$-moduli space. Hence, we can restrict ourselves to four special cases: 
\begin{itemize}
\item[i)] the generic $\mathbbm T^2/\Z{2}$ orbifold with generic vev $\langle U \rangle$, 
\item[ii)] the tetrahedron with $\vev U=e^{\nicefrac{\pi\I}{3}}$,
\item[iii)] the raviolo with $\vev U=\I$, and
\item[iv)] $\mathbbm T^2/\Z{2}$ orbifolds with \CP-enhancement.
\end{itemize}
In the following, we will discuss these four cases in detail, restricting to points of the 
fundamental domain in moduli space.

%%%%%%%%%%%%%%%%%%%%%%%%%%%%%%%%%%%%%%%%%%%%%%%%%%%%%%%%%%%%%%%%%%%%%%%
\subsection[The generic T2/Z2 orbifold]{\boldmath The generic $\mathbbm{T}^2/\Z{2}$ orbifold\unboldmath}

\begin{figure}[t!]
    \centering
    \begin{minipage}{0.48\textwidth}%
        \subfloat[(a)][$U$ fixed at a generic value $\vev U\neq\I,e^{\nicefrac{\pi\I}{3}}$.]{
            \label{fig:genericU}
            \centering \includegraphics[page=1,scale=1.8]{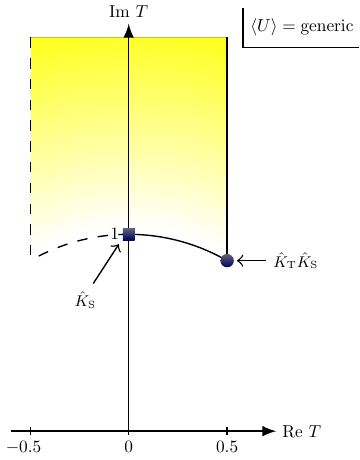}
            }
    \end{minipage}%
    \begin{minipage}{0.48\textwidth}%
        \subfloat[(b)][$T$ fixed at a generic value $\vev T\neq\I,e^{\nicefrac{\pi\I}{3}}$.]{
            \label{fig:genericT}
            \centering \includegraphics[page=1,scale=1.8]{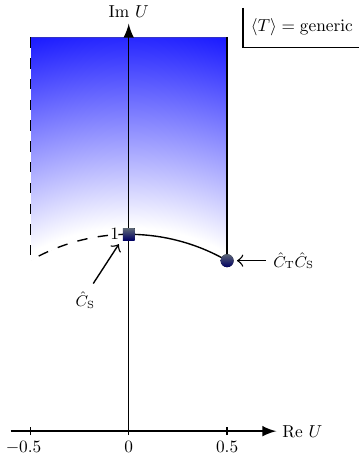}
        }
    \end{minipage}%
    \caption{Generators of nontrivial stabilizer subgroups at special points in moduli space. 
    (a) For generic $\vev U$, only at $\vev T=\I$ 
    (square) and $\vev T=e^{\nicefrac{\pi\I}{3}}$ (bullet) the stabilizer subgroup is nontrivial.
    The corresponding stabilizer subgroups are $H_{(\I,\vev U)}=\langle \hat{K}_\mathrm{S}\rangle\cong\Z2$ and 
    $H_{(e^{\nicefrac{\pi\I}{3}},\vev U)}=\langle \hat{K}_\mathrm{T} \hat{K}_\mathrm{S}\rangle\cong\Z3$.
    (b) For generic $\vev T$, the results are equivalent due to mirror symmetry $\hat M$, which
    exchanges $T\leftrightarrow U$, $\hat{K}_\mathrm{T}\leftrightarrow\hat{C}_\mathrm{T}$ and
    $\hat{K}_\mathrm{S}\leftrightarrow\hat{C}_\mathrm{S}$.
    \label{UTmodulispaces}}
\end{figure}

Let us consider first the case of a generic vev $\vev U$ of the complex structure modulus $U$. In 
this case, \CP is broken by the generic vev of $U$. Furthermore, as shown in 
figure~\ref{fig:genericU}, there are only two inequivalent special values of the K\"ahler modulus 
$T$ associated with a nontrivial stabilizer modular subgroup: $\vev T=\I$ and 
$\vev T= e^{\nicefrac{\pi\I}{3}}$. At these points, we see that the K\"ahler modulus is invariant 
under the transformations
\begin{subequations}
\label{eq:StabilizerGeneratorsT}
\begin{eqnarray}
  \text{at } \vev T = \I \hspace{5.5mm}&:&\vev T ~\stackrel{\hat{K}_\mathrm{S}}{\longrightarrow}~ -\frac{1}{\vev T} ~=~ \vev T\;,\\
  \text{at } \vev T = e^{\nicefrac{\pi\I}{3}}&:&\vev T ~\stackrel{\hat{K}_\mathrm{T}}{\longrightarrow}~ \vev T+1 ~\stackrel{\hat{K}_\mathrm{S}}{\longrightarrow}~ -\frac{1}{\vev T}+1 ~=~ \vev T\;,
\end{eqnarray}
\end{subequations}
and $\vev U$ is not affected, cf.~eq.~\eqref{eq:SLZ2UandT}. The corresponding stabilizer subgroups are
\begin{subequations}
\label{eq:StabilizerSubgroupsT}
\begin{eqnarray}
\label{eq:StabilizerSubgroupsTKS}
  \text{at } \vev T = \I \hspace{5.5mm}&:& H_{(\I,\vev U)} \hspace{6.5mm}=~  \big\langle~ \hat{K}_\mathrm{S} ~|~ (\hat{K}_\mathrm{S})^2\sim\Id~\big\rangle\hspace{1.255cm}\cong~\Z2\;,\\
\label{eq:StabilizerSubgroupsTKTKS}
  \text{at } \vev T = e^{\nicefrac{\pi\I}{3}} &:& H_{(e^{\nicefrac{\pi\I}{3}},\vev U)} ~=~  \big\langle~\hat{K}_\mathrm{T}\hat{K}_\mathrm{S} ~|~ (\hat{K}_\mathrm{T}\hat{K}_\mathrm{S})^3=\Id~\big\rangle ~\cong~ \Z3\;.
\end{eqnarray}
\end{subequations}
Note that $(\hat{K}_\mathrm{S})^2$ is a trivial element of all stabilizer subgroups because the 
stabilizer subgroup is defined up to point group transformations (see the discussion around 
eq.~\eqref{eq:stabilizerDefinition}) and $(\hat{K}_\mathrm{S})^2=(\hat{C}_\mathrm{S})^2=-\Id_4$ 
is equivalent to the \Z2 point group generator $\hat\Theta=-\Id_4$. It is easy to confirm the 
mirror duals of eqs.~\eqref{eq:StabilizerSubgroupsT}, as shown in figure~\ref{fig:genericT} for 
generic \vev{T}.

To exemplify the enhancement of the traditional flavor symmetry to unified flavor groups, let us 
focus on the hypersurface $\vev T=e^{\nicefrac{\pi\I}{3}}$ for generic $\vev U$. We use the 
$\rho_{\rep{4}}$ representation of $\hat{K}_\mathrm{T}$ and $\hat{K}_\mathrm{S}$ given in 
eqs.~\eqref{eq:SL2ZTOfTwistedStrings}, where $(\rho_{\rep{4}}(\hat{K}_\mathrm{S}))^2=\Id_4$, 
together with those of the generators of the traditional flavor group, 
eqs.~\eqref{eq:OuterOfTwistedStrings} and eqs.~\eqref{eq:SelectionRulesOfTwistedStrings}. In this 
way, we find that the nontrivial stabilizer subgroup enhances
\begin{equation}
\label{eq:UnifiedFlavorAtT=exp(pii/3)}
  (D_8\x D_8)/\Z2 \quad\mathrm{to}\quad [96,204]\qquad\mathrm{at}\quad \vev U=\mathrm{generic},\;\vev T = e^{\nicefrac{\pi\I}{3}}\;.
\end{equation}
As intuitively expected, the order is 3 times as large as the order of the original group 
$(D_8\x D_8)/\Z2 \cong [32, 49]$ because of the nontrivial \Z3 factor introduced by the 
stabilizer. Due to mirror symmetry, this unified flavor symmetry is isomorphic to the resulting 
unified flavor symmetry at $\vev U= e^{\nicefrac{\pi\I}{3}}$ for generic $\vev T$.

After this first case, we can now proceed to study the more complex cases of the $\mathbbm T^2/\Z2$ orbifold
adopting the shapes of a tetrahedron and a raviolo, where the \CP-like modular transformation $\hat\Sigma_*$
plays an important role.

%%%%%%%%%%%%%%%%%%%%%%%%%%%%%%%%%%%%%%%%%%%%%%%%%%%%%%%%%%%%%%%%%%%%%%%
\subsection[The tetrahedron with U=exp(pi i/3)]{\boldmath The tetrahedron with $\vev U=e^{\nicefrac{\pi\I}{3}}$\unboldmath}

\begin{figure*}[t!]
\centering{\includegraphics[width=0.94\linewidth]{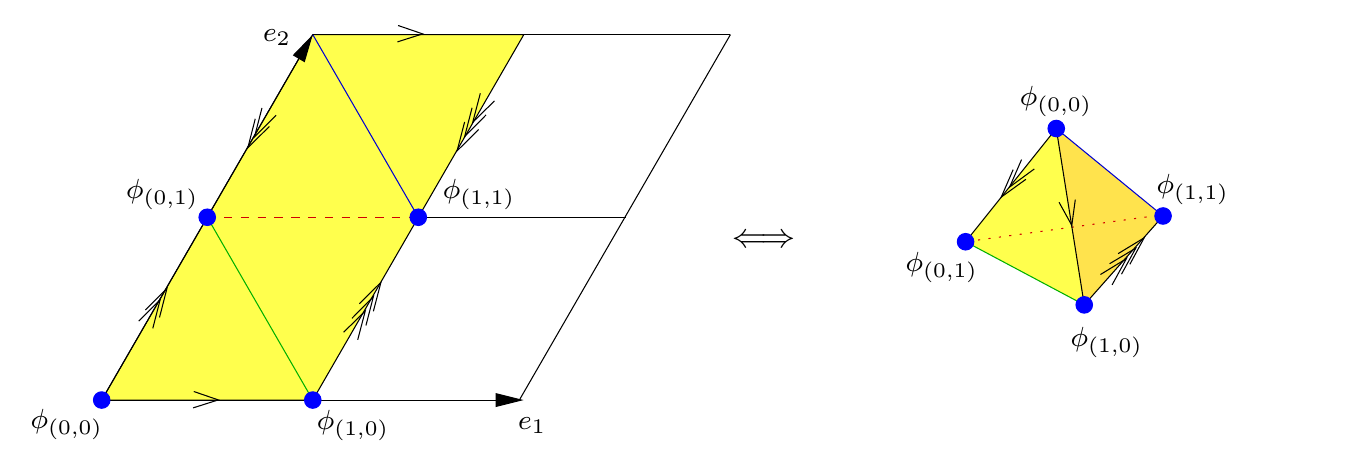}}
\caption{The $\mathbbm T^2/\Z2$ orbifold with $\vev U=e^{\nicefrac{\pi\I}{3}}$. Identifying the arrows 
indicated on the boundaries of the fundamental domain of the orbifold yields a tetrahedron, whose vertex
corners correspond to the four fixed points of the orbifold, where twisted matter fields $\phi_{(n_1,n_2)}$ 
are localized.}
\label{fig:Z2Tetrahedron}
\end{figure*}

Setting $\vev U=e^{\nicefrac{\pi\I}{3}}$ leads to the $\mathbbm T^2/\Z2$ illustrated in figure~\ref{fig:Z2Tetrahedron}.
On the left, we see the corresponding two-dimensional torus lattice spanned by $e_1$ and $e_2$ of equal length
and enclosing an angle of $\nicefrac{\pi}{3}$.
Modding out a \Z2 symmetry of the two-torus reduces the fundamental domain to the yellow region, whose 
boundaries are identified as the arrows indicate. These identifications allow for the fundamental domain of 
the $\mathbbm T^2/\Z2$ orbifold to adopt the shape of the tetrahedron displayed on the right of the figure.
The corners of the tetrahedron correspond to the four fixed points of the orbifold, where the twisted matter
fields $(\phi_{(0,0)}, \phi_{(1,0)}, \phi_{(0,1)}, \phi_{(1,1)})^\mathrm{T}$ are localized.

The generic flavor symmetry group of the tetrahedron can be found by considering the stabilizer 
subgroup at $\vev U=e^{\nicefrac{\pi\I}{3}}$ for generic $\vev T$, 
\begin{equation}
H_{(\vev T, e^{\nicefrac{\pi\I}{3}})} ~=~ \big\langle~ \hat{C}_\mathrm{T}\hat{C}_\mathrm{S}~\big\rangle ~\cong~ \Z{3}\;,
\end{equation}
which corresponds to the mirror dual of eq.~\eqref{eq:StabilizerSubgroupsTKTKS}. In this case, the 
unified flavor group is constructed by the generators of the traditional flavor group and 
$\hat{C}_\mathrm{T}\hat{C}_\mathrm{S}$. The resulting unified flavor group is $[96,204]$, as 
in the dual scenario given in eq.~\eqref{eq:UnifiedFlavorAtT=exp(pii/3)}.

\begin{figure}[t!]
    \centering
    \begin{minipage}{0.48\textwidth}%
        \subfloat[b][$\mathbbm T^2/\Z2$ with $\vev U=e^{\nicefrac{\pi\I}{3}}$]{
            \label{fig:TplaneForU=e^pii/3}
            \centering \includegraphics[page=3,scale=1.8]{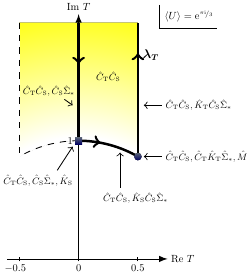}
        }
    \end{minipage}%
    \begin{minipage}{0.48\textwidth}%
        \subfloat[a][$\mathbbm T^2/\Z2$ with $\vev U=\I$]{
            \label{fig:TplaneForU=i}
            \centering \includegraphics[page=2,scale=1.8]{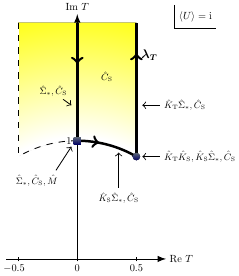}
        }
    \end{minipage}%
    \caption{Generators of the stabilizer subgroups $H_{(\vev T,\vev U)}$ at different points \vev{T} of the fundamental 
    domain of $\SL{2,\Z{}}_T$ for two special vevs \vev{U} of the complex structure modulus. These elements 
    enhance the traditional flavor symmetry to various unified flavor groups.\label{fig:TplaneRavioloTetrahedron}}
\end{figure}

Since the stabilizer generator $\hat{C}_\mathrm{T}\hat{C}_\mathrm{S}$ is now a symmetry everywhere 
in K\"ahler moduli space, it is displayed in the fundamental domain (yellow area) of 
figure~\ref{fig:TplaneForU=e^pii/3}. If we now consider the \CP-like transformation $\hat\Sigma_*$, 
for particular values of $\vev T$, the flavor group of the tetrahedron is enhanced further. These 
enhancements occur at the points \vev{T} along the curve $\lambda_T$ of 
figure~\ref{fig:TplaneForU=e^pii/3}, which is the boundary of the fundamental domain in $T$ moduli 
space. The orientation of the curve $\lambda_T$ is indicated by bold arrows. It will be used 
later in section~\ref{sec:CPMoudliSpace}.

For example, at the points of the curve $\lambda_T$ with $\mathrm{Re}\vev{T}=\nicefrac12$ and 
$\mathrm{Im}\vev{T}>\nicefrac{\sqrt3}{2}$ we find the \CP-like transformation 
$\hat{K}_\mathrm{T}\hat{C}_\mathrm{S}\hat\Sigma_*$. It acts on the moduli as
\begin{equation}
\label{eq:StabilizerGeneratorsForT=1/2U=exp(pii/3)}
  \begin{array}{cccccccl}
    \vev T & \stackrel{\hat{K}_\mathrm{T}}{\longrightarrow}& \vev T+1 &\stackrel{\hat{C}_\mathrm{S}}{\longrightarrow}&\vev T+1 
                                &\stackrel{\hat\Sigma_*}{\longrightarrow}& -\vev{\bar{T}}+1 &=~ \vev T\;, \\
    \vev U & \stackrel{\hat{K}_\mathrm{T}}{\longrightarrow}& \vev U   &\stackrel{\hat{C}_\mathrm{S}}{\longrightarrow}&-\dfrac{1}{\vev U} 
                                &\stackrel{\hat\Sigma_*}{\longrightarrow}& \dfrac{1}{\vev{\bar{U}}} &=~ \vev U\;.
  \end{array}
\end{equation}
Hence, it belongs to the stabilizer subgroup
\begin{equation}
H_{(\nicefrac12+\I\, \text{Im}\vev{T},\, e^{\nicefrac{\pi\I}{3}})}~=~ \big\langle~ \hat{C}_\mathrm{T}\hat{C}_\mathrm{S},\, \hat{K}_\mathrm{T}\hat{C}_\mathrm{S}\hat\Sigma_*~\big\rangle\;,
\qquad\text{with }\text{Im}\vev{T}>\nicefrac{\sqrt3}{2}\;.
\end{equation}
Using the representations of the involved modular transformations, given in 
eqs.~\eqref{eq:SL2ZUOfTwistedStrings},~\eqref{eq:SL2ZTOfTwistedStrings} and~\eqref{eq:SigmaOnTwistedStrings}, 
leads to the unified flavor group $[192,1494]$, which is known as $SW_4$ and can be associated with 
all pure rotations of a four-dimensional cube~\cite{Baake:1981qe}. In contrast to 
ref.~\cite{Kobayashi:2006wq}, we realize the $SW_4$ symmetry only if both moduli $T$ and $U$ take 
special values. The same enhancement results for $\vev T = e^{\I\varphi}$ with 
$\nicefrac{\pi}{3} < \varphi < \nicefrac{\pi}{2}$, where, as indicated in figure~\ref{fig:TplaneForU=e^pii/3},
the stabilizer subgroup $H_{(e^{\I\varphi},e^{\nicefrac{\pi\I}{3}})}$ is generated by $\hat{C}_\mathrm{T}\hat{C}_\mathrm{S}$ 
and $\hat{K}_\mathrm{S}\hat{C}_\mathrm{S}\hat\Sigma_*$.

Another interesting example is the maximally symmetric point $\vev T=\vev U=e^{\nicefrac{\pi\I}{3}}$.
There, in addition to $\hat{C}_\mathrm{T}\hat{C}_\mathrm{S}$, also the mirror transformation
$\hat M$ and the \CP-like transformation
\begin{equation}    
\label{eq:StabilizerGeneratorsForT=U=exp(pii/3)}
  \begin{array}{cccccccl}
    \vev T & \stackrel{\hat{C}_\mathrm{T}}{\longrightarrow}& \vev T   &\stackrel{\hat{K}_\mathrm{T}}{\longrightarrow}&\vev T+1 
                                &\stackrel{\hat\Sigma_*}{\longrightarrow}& -\vev{\bar{T}}+1 &=~ \vev T\;, \\
    \vev U & \stackrel{\hat{C}_\mathrm{T}}{\longrightarrow}& \vev U+1 &\stackrel{\hat{K}_\mathrm{T}}{\longrightarrow}&\vev U+1
                                &\stackrel{\hat\Sigma_*}{\longrightarrow}& -\vev{\bar{U}}+1 &=~ \vev U
  \end{array}
\end{equation}
build the stabilizer subgroup
\begin{equation}
H_{(e^{\nicefrac{\pi\I}{3}},e^{\nicefrac{\pi\I}{3}})}~=~ \big\langle~ \hat{C}_\mathrm{T}\hat{C}_\mathrm{S},\,\hat{C}_\mathrm{T}\hat{K}_\mathrm{T}\hat\Sigma_*,\,\hat M~\big\rangle\;,
\end{equation}
as displayed in figure~\ref{fig:TplaneForU=e^pii/3}. Using the representations of these generators 
acting on twisted matter fields, including eq.~\eqref{eq:repM}, we find that the generic flavor 
symmetry of the tetrahedron $[96,204]$ is enhanced to the unified flavor group $[1152,157463]$. 
This corresponds to the largest enhancement of the traditional flavor group in the 
$\mathbbm T^2/\Z2$ orbifold, including mirror symmetry and \CP. We observe that even in the 
simplest case of a $\mathbbm T^2/\Z2$ orbifold, the flavor symmetry can be very large.

Other nontrivial stabilizer subgroup generators are displayed along the curve $\lambda_T$ of 
figure~\ref{fig:TplaneForU=e^pii/3}.

%%%%%%%%%%%%%%%%%%%%%%%%%%%%%%%%%%%%%%%%%%%%%%%%%%%%%%%%%%%%%%%%%%%%%%%
\subsection[The raviolo with <U>=i]{\boldmath The raviolo with $\vev U=\I$\unboldmath}

Fixing the complex structure modulus to $\vev U=\I$ amounts to setting $|e_1|=|e_2|$ with an angle 
of $\nicefrac{\pi}{2}$ between them. This corresponds to a toroidal lattice whose basis vectors are 
orthogonal and have equal length. In this case, the $\mathbbm T^2/\Z2$ orbifold takes the 
shape of a raviolo, similar to the one depicted in figure~\ref{fig:Z2Orbifold}, but whose edges are 
perpendicular and all have the same length.

The stabilizer subgroup for generic \vev{T} and $\vev U=\I$ is dual to the one given in
eq.~\eqref{eq:StabilizerSubgroupsTKS}: it is generated by $\hat{C}_\mathrm{S}$. Note that at 
$\vev U=\I$ the transformation $\hat C_\mathrm{S}$ gives rise to a $\nicefrac{\pi}{2}$ rotation in 
the compact dimensions. Hence, $\hat C_\mathrm{S}$ is a discrete remnant of the 
higher-dimensional Lorentz symmetry. Thus, it generates an $R$-symmetry. In figure~\ref{fig:TplaneForU=i}, 
this universal generator of the stabilizer subgroups is displayed in the fundamental domain in the 
$T$ moduli space (yellow area). Similarly to the previous case, taking into account the 
\CP-like transformation $\hat\Sigma_*$, further enhancements arise from the elements of the
stabilizer subgroups shown in figure~\ref{fig:TplaneForU=i} at the points \vev{T} along the curve 
$\lambda_T$.

%%%%%%%%%%%%%%%%%%%%%%%%%%%%%%%%%%%%%%%%%%%%%%%%%%%%%%%%%%%%%%%%%%%%%%%
\subsection[Z2 orbifolds with CP-enhancement]{\boldmath $\mathbbm T^2/\Z{2}$ orbifolds with \CP-enhancement\unboldmath}
\label{sec:CPMoudliSpace}

\begin{figure}[t!]
\centering
\centering{\includegraphics[width=0.9\linewidth]{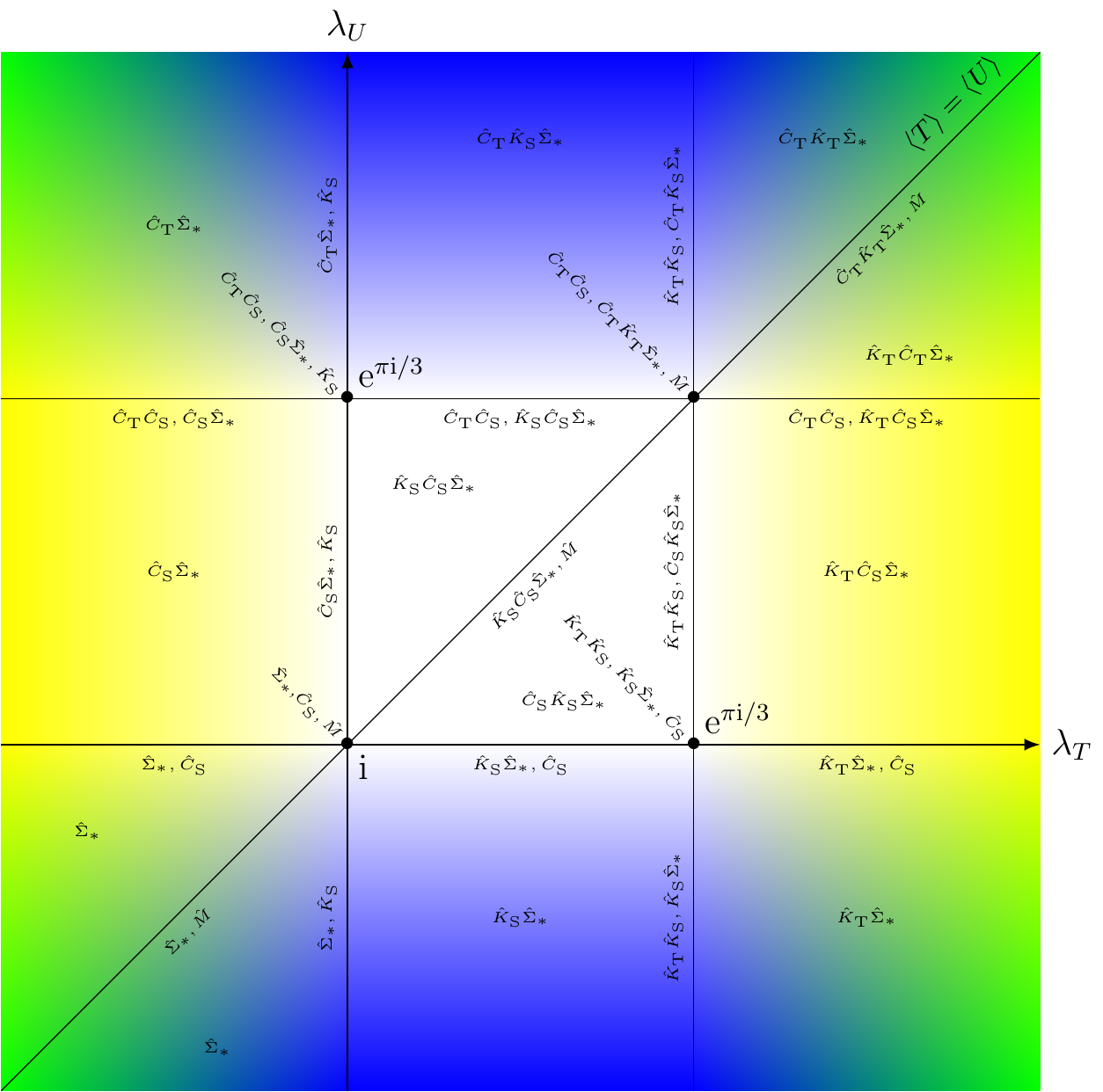}}
\caption{Generators of the stabilizers $H_{(\vev{T},\vev{U})}$ for $\vev{T}\in\lambda_T$ and 
$\vev{U}\in\lambda_U$. The axes $\lambda_T$ and $\lambda_U$ correspond to the curves on the 
boundaries of the two-dimensional fundamental domains of $T$ and $U$, see e.g.\
figure~\ref{fig:TplaneRavioloTetrahedron} for $\lambda_T$. The diagonal depicts the hypersurface 
where $\vev U=\vev T$ on the curves $\lambda_T$ and $\lambda_U$. The stabilizers above and below 
the diagonal are related by mirror symmetry $\hat M$.\label{fig:LocalFlavorGroups}}
\end{figure}

In figure~\ref{fig:LocalFlavorGroups}, we display the landscape of all nontrivial stabilizers that 
contain $\CP$-like generators. To do so, we use the straightened curves $\lambda_T$ and $\lambda_U$ 
as the axes in figure~\ref{fig:LocalFlavorGroups}. To be specific, the curve $\lambda_T$ is defined 
as the boundary of the fundamental domain of the $S_3^T$ modular symmetry in 
figure~\ref{fig:TplaneRavioloTetrahedron}, where the orientation is illustrated by the bold arrows. 
Its mirror dual $\lambda_U$ is defined analogously in the $U$-modulus plane. Hence, the horizontal 
$\lambda_T$ axis must be interpreted as follows:
\vspace{-2mm}
\begin{itemize}
\item between the leftmost point and $\lambda_T=\I$, $\lambda_T$ corresponds to 
$\infty>\mathrm{Im}\vev{T}\geq1$ and $\mathrm{Re}\vev T=0$,
\item for $\lambda_T$ between $\I$ and $e^{\nicefrac{\pi\I}{3}}$, the curve $\lambda_T$ describes 
$\vev T=e^{\I\varphi}$ with $\nicefrac{\pi}{2}>\varphi>\nicefrac{\pi}{3}$, and
\item between $\lambda_T=e^{\nicefrac{\pi\I}{3}}$ and the rightmost point, the curve $\lambda_T$ is 
associated with $\nicefrac{\sqrt3}{2}\leq\text{Im}\vev{T}<\infty$ and $\text{Re}\vev T=\nicefrac12$.
\end{itemize}
Similarly, the dual vertical axis $\lambda_U$ has to be read bottom-up, exchanging $\vev T$ in the 
previous description by $\vev U$. The color schema is such that points with values closer to 
$\vev T\to\I\infty$ ($\vev U\to\I\infty$) are more yellow (blue). Large imaginary values of both 
$\vev T$ and $\vev U$ yield the green texture.

We observe that the results presented in figures~\ref{fig:TplaneForU=e^pii/3} and~\ref{fig:TplaneForU=i}
are reproduced in figure~\ref{fig:LocalFlavorGroups} along the upper and lower horizontal lines, respectively.
Mirror symmetry maps the lower (upper) horizontal line to the left (right) vertical line, where the stabilizer generators
coincide after performing the transformations $\hat C_\mathrm{S}\stackrel{\hat M}{\longleftrightarrow}\hat K_\mathrm{S}$
and $\hat C_\mathrm{T}\stackrel{\hat M}{\longleftrightarrow}\hat K_\mathrm{T}$. 

The diagonal in figure~\ref{fig:LocalFlavorGroups} describes all special points in moduli space at 
which $\vev T = \vev U$ is satisfied. We note that there are different enhancements depending on 
the particular values of the moduli. For example, for $\vev T = \vev U = \I x$ with $x>1$, the 
stabilizer subgroup is $H_{(\I x,\I x)} = \langle \hat\Sigma_*, \hat M\rangle$, as displayed on the 
lower left panel of the figure.
As a final example, consider the area $\lambda_T<\I$ and $\I<\lambda_U< e^{\nicefrac{\pi\I}{3}}$ 
in figure~\ref{fig:LocalFlavorGroups}, i.e.\ the middle left panel (equivalently defined by $\vev T=\I x$ 
with $x>1$ and $\vev U=e^{\I\varphi}$ for $\nicefrac{\pi}{2}>\varphi>\nicefrac{\pi}{3}$). In this case, 
the stabilizer subgroup is generated by the \CP-like transformation $\hat C_\mathrm{S}\hat\Sigma_*$.
In summary, we obtain all nontrivial stabilizers that involve \CP-enhancement and illustrate 
them in figure~\ref{fig:LocalFlavorGroups}.

%%%%%%%%%%%%%%%%%%%%%%%%%%%%%%%%%%%%%%%%%%%%%%%%%%%%%%%%%%%%%%%%%%%%%%%%%%%%%%%%%%%%%%%%%%%%%%%%%%%%%%%%%%%%%%%%%%%%%%%%%%%%%%%%%%%%%%%%%%
\section{Conclusions and Outlook}
\label{sec:conclusions}

We have seen that modular transformations of the $\mathbbm T^2/\Z2$ orbifold lead to an exceedingly 
rich eclectic structure. From the traditional flavor symmetry $D_8\times D_8/\Z2\cong [32,49]$, we 
obtain an eclectic flavor group with as many as 4608 elements. We also observe large flavor groups 
that are linearly realized at specific regions in $T$- and $U$-moduli space, the largest being 
$[1152,157463]$ at $\vev T=\vev U=\exp(\nicefrac{\pi\I}{3})$. For reasons of clarity and 
simplicity, the discussion in the present paper has concentrated on group theoretical studies of 
finite modular groups and we have not yet included a full implementation of the automorphy factors 
of $\SL{2,\Z{}}_T$ and $\SL{2,\Z{}}_U$. Future work~\cite{Baur:2020pr} will include this along the 
lines explained in refs.~\cite{Nilles:2020tdp,Nilles:2020pr} for the $\mathbbm T^2/\Z3$ orbifold. 
As discussed in ref.~\cite{Nilles:2020tdp}, the automorphy factors lead to discrete phases. These 
give rise to discrete $R$-symmetries that complete the full eclectic structure. 
%Together with the 
%results presented in the present paper, this will then be the last crucial step towards a 
%discussion of orbifolds with six compact extra dimensions and allow the analysis of flavor 
%symmetries of realistic low energy theories of particle physics derived from string theory.
Together with the results presented in this paper, this will be a crucial step towards a discussion 
of orbifolds with six compact extra dimensions. The general case in $D=6$ will be too difficult to 
be analyzed in detail. Using our $D=2$ building blocks we can simplify the discussion and consider 
elliptic fibrations of the $D=6$ case with various $D=2$ sublattice rotations. In a further step we 
would then have to consider specific string models that could successfully describe the flavor 
structure of quarks and leptons and make contact with available botton-up 
constructions~\cite{Feruglio:2019ktm}. The predictions of the model will crucially depend on the 
value of the moduli that potentially break flavor and \CP in a desirable way. Thus, the ultimate 
step in realistic model building will be a discussion of moduli stabilization. We think it is 
premature to discuss this mechanism at the moment, as the process of moduli stabilization will 
crucially depend on the specific models under consideration. So we rather first would like to 
construct models that have a chance for a realistic description of flavor for some values of the 
moduli and relegate the discussion of moduli stabilization to future work.

\section*{Acknowledgments}
We thank Michael Ratz for useful discussions. 
A.B., S.R.-S.\ and P.V.\ are supported by the Deutsche Forschungsgemeinschaft (SFB1258).
The work of S.R.-S.\ was partly supported by CONACyT grants F-252167 and 278017.

%%%%%%%%%%%%%%%%%%%%%%%%%%%%%%%%%%%%%%%%%%%%%%%%%%%%%%%%%%%%%%%%%%%%%%%%%%%%%%%%%%%%%%%%%%%%%%%%%%%%%%%%%%%%%%%%%%%%%%%%%%%%%%%%%%%%%%%%%%
\appendix

\vspace{-2mm}
\section{Narain lattice}
\label{app:NarainLattice}

In this appendix, we give a brief discussion on the Narain space group of our $\mathbbm{T}^2/\Z{2}$ 
orbifold, its outer automorphisms and their actions on the twisted matter fields that are 
localized at the fixed points of the $\mathbbm{T}^2/\Z{2}$ orbifold.

\subsection{Orbifolds defined by the Narain space group}

The string $y(\tau,\sigma)$ can be split into right- and left-moving degrees of freedom 
$y_\mathrm{R}$ and $y_\mathrm{L}$, respectively, where
\begin{equation}
\begin{pmatrix}y\\\tilde{y}\end{pmatrix}~=~\frac{1}{\sqrt{2}}\begin{pmatrix}\Id_2&\Id_2\\-\Id_2&\Id_2\end{pmatrix}\,\begin{pmatrix}y_\mathrm{R}\\y_\mathrm{L}\end{pmatrix}\;,
\end{equation}
and $\tilde{y}$ denotes the so-called dual string coordinate. Then, one defines Narain coordinates 
$Y$ and compactifies them on a Narain torus, i.e.
\begin{equation}
Y ~:=~ \left(\begin{array}{c}y_\mathrm{R}\\y_\mathrm{L}\end{array}\right) \quad\mathrm{where}\quad Y ~\sim~ Y + E\,N \quad\mathrm{and}\quad E\,N ~\in~\Gamma\;,
\end{equation}
Here, the integer vector $N = (n_1,n_2,m_1,m_2)^\mathrm{T}\in\Z{}^4$ gives the winding numbers 
$(n_1,n_2)$ and the Kaluza--Klein numbers $(m_1,m_2)$. In addition, $E$ is a $4 \times 4$ vielbein 
matrix, the so-called Narain vielbein (given for example in refs.~\cite{Baur:2019kwi,Baur:2019iai}, changing $B$ to $-B$). 
To ensure two-dimensional worldsheet modular invariance, the Narain vielbein has to span an even, 
integer, self-dual lattice with metric $\eta=\mathrm{diag}(-1,-1,1,1)$ of signature $(2,2)$: the 
so-called Narain lattice $\Gamma$. In a second step, one can mod out the Narain lattice by a 
$\Z{K}$ rotational symmetry, generated by the Narain twist
\begin{equation}
\Theta ~=~ \begin{pmatrix}\theta_\mathrm{R} & 0\\0 & \theta_\mathrm{L}\end{pmatrix} \qquad\mathrm{where}\qquad \Theta^K ~=~ \Id_4 \quad\mathrm{and}\quad \Theta\Gamma~=~\Gamma\;.
\end{equation}
Hence, one defines an orbifold in the Narain formulation of string theory as
\begin{equation}
Y ~\mapsto~ g\,Y := \Theta^k\,Y + E\,N ~\sim~ Y\;,
\end{equation}
where $g = (\Theta^k, E\,N)$ is an element of the so-called Narain space group $S_\mathrm{Narain}$. 
The orbifold is called symmetric if the Narain twist $\Theta$ acts identically on right- and 
left-movers, i.e.\ $\theta :=\theta_\mathrm{R}=\theta_\mathrm{L}$.

\subsection[Outer automorphisms of the Z2 Narain space group]{\boldmath Outer automorphisms of the $\Z{2}$ Narain space group\unboldmath}
\label{app:Out}

Discrete symmetries of the four-dimensional effective theory obtained from orbifold 
compactifications are associated with the outer automorphisms of the Narain space group 
$S_\mathrm{Narain}$. In analogy to the definition in eq.~\eqref{eq:OuterOfS} for the geometrical 
space group, an outer automorphism of $S_\mathrm{Narain}$ is defined as a mapping from 
$S_\mathrm{Narain}$ to itself. In more detail, a transformation 
$h=\big(\Sigma,E\,T\big)\not\in S_\mathrm{Narain}$ is an outer automorphism of $S_\mathrm{Narain}$ 
if
\begin{equation}\label{eq:OuterOfSNarain}
g ~\stackrel{h}{\longmapsto}~ h^{-1}\,g\,h ~\stackrel{!}{\in}~ S_\mathrm{Narain}\;,
\end{equation}
for all $g = \big(\Theta^k, E\,N\big) \in S_\mathrm{Narain}$. In addition, $\Sigma$ has to preserve 
the Narain metric
\begin{equation}\label{eq:Oeta}
\Sigma^\mathrm{T}\eta\,\Sigma ~=~ \eta\;.
\end{equation}

Next, we translate the Narain space group and its outer automorphisms into the lattice basis, in 
which we denote all quantities in general by a hat. For example, eq.~\eqref{eq:OuterOfSNarain} 
reads in the lattice basis
\begin{equation}\label{eq:OuterOfSNarainLatticeBasis}
\hat{g} ~\stackrel{\hat{h}}{\longmapsto}~ \hat{h}^{-1}\,\hat{g}\,\hat{h} ~\stackrel{!}{\in}~ \hat{S}_\mathrm{Narain}\;,
\end{equation}
where
\begin{subequations}
\begin{eqnarray}
\hat{g} & := & \big(E^{-1},0\big)\, \big(\Theta^k, E\,N\big)\, \big(E,0\big) ~=~ \big(\hat\Theta^k, N\big) ~\in~     \hat{S}_\mathrm{Narain} \quad\mathrm{and}\\
\hat{h} & := & \big(E^{-1},0\big)\, \big(\Sigma, E\,T\big)\,   \big(E,0\big) ~=~ \big(\hat\Sigma, T\big)   ~\not\in~ \hat{S}_\mathrm{Narain}\;.
\end{eqnarray}
\end{subequations}
Here, we have defined $\hat\Theta:=E^{-1}\Theta\,E$ and $\hat\Sigma:=E^{-1}\Sigma\,E$. Furthermore, 
due to eq.~\eqref{eq:Oeta} we have to impose $\hat\Sigma \in \mathrm{O}_{\hat\eta}(2,2,\Z{})$, where the 
group $\mathrm{O}_{\hat\eta}(2,2,\Z{})$ of ``rotational'' outer automorphisms of the Narain lattice $\Gamma$ 
is defined as
\begin{equation}\label{eq:OEtaHat22Z}
\mathrm{O}_{\hat\eta}(2,2,\Z{}) ~:=~ \big\langle ~\hat\Sigma~\big|~ \hat\Sigma ~\in~\mathrm{GL}(4,\Z{}) \quad\mathrm{with}\quad \hat\Sigma^\mathrm{T}\hat\eta\,\hat\Sigma = \hat\eta~\big\rangle\;,
\end{equation}
using the Narain metric in the lattice basis
\begin{equation}\label{eq:NarainMetric}
\hat\eta ~:=~ E^\mathrm{T}\eta\,E ~=~ \begin{pmatrix}0 & \Id_2\\\Id_2 & 0\end{pmatrix}\;.
\end{equation}
As discussed in detail in ref.~\cite{Baur:2019iai}, the group $\mathrm{O}_{\hat\eta}(2,2,\Z{})$ 
contains the generators
\begin{subequations}\label{eq:ModularGroupOfT2Part1}
\begin{eqnarray}
\hat{K}_\mathrm{S}~:=~\begin{pmatrix}0&\epsilon\\ \epsilon&0\end{pmatrix} &\mathrm{and}& \hat{K}_\mathrm{T}~:=~\begin{pmatrix}\Id_2&0\\ \epsilon&\Id_2\end{pmatrix}\;,\quad\mathrm{with}\quad\epsilon~:=~\begin{pmatrix}0& 1\\ -1 & 0\end{pmatrix}\;,\\
\hat{C}_\mathrm{S}~:=~\begin{pmatrix}-\epsilon&0\\0&-\epsilon\end{pmatrix} &\mathrm{and}& \hat{C}_\mathrm{T}~:=~\begin{pmatrix}\gamma&0\\0&\gamma^{-T} \end{pmatrix}\;,\quad\mathrm{with}\quad\gamma  ~:=~\begin{pmatrix}1&-1\\  0 & 1\end{pmatrix}\;.
\end{eqnarray}
\end{subequations}
Note that, compared to refs.~\cite{Baur:2019iai,Nilles:2020kgo}, we redefined $\hat{K}_\mathrm{S}$, 
$\hat{K}_\mathrm{T}$ and $\hat{C}_\mathrm{S}$. $\hat{K}_\mathrm{S}$ 
and $\hat{K}_\mathrm{T}$ generate $\SL{2,\Z{}}_T$ of the K\"ahler modulus $T$, while 
$\hat{C}_\mathrm{S}$ and $\hat{C}_\mathrm{T}$ generate $\SL{2,\Z{}}_U$ of the complex structure 
modulus $U$. Moreover, the group $\mathrm{O}_{\hat\eta}(2,2,\Z{})$ contains two additional 
generators, given by
\begin{equation}\label{eq:ModularGroupOfT2Part2}
\hat\Sigma_* ~:=~ \begin{pmatrix}-1&0&0&0\\0&1&0&0\\0&0&-1&0\\0&0&0&1\end{pmatrix} \quad\mathrm{and}\quad \hat{M} ~:=~ \begin{pmatrix}0&0&-1&0\\0&1&0&0\\ -1&0&0&0\\0&0&0&1\end{pmatrix}\;.
\end{equation}
The first generator $\hat\Sigma_*$ gives rise to a $\CP$-like transformation, while $\hat{M}$ 
generates a mirror symmetry, i.e. $T \stackrel{\hat{M}}{\longleftrightarrow} U$ and
\begin{equation}\label{eq:mirrorsymmetry}
\hat{M}\,\hat{C}_\mathrm{S}\,\hat{M}^{-1} ~=~ \hat{K}_\mathrm{S} \quad\mathrm{and}\quad \hat{M}\,\hat{C}_\mathrm{T}\,\hat{M}^{-1} ~=~ \hat{K}_\mathrm{T}\;.
\end{equation}

In order to identify the transformation of the moduli $T$ and $U$ under $\mathrm{O}_{\hat\eta}(2,2,\Z{})$, 
one can consider the generalized metric $\mathcal{H}$, defined by
\begin{equation}
\mathcal{H}(T,U) ~:=~ E^\mathrm{T}E\;.
\end{equation}
Under a modular transformation the Narain vielbein transforms as $E \mapsto E\,\hat\Sigma^{-1}$ for 
$\hat\Sigma \in \mathrm{O}_{\hat\eta}(2,2,\Z{})$. Consequently, we obtain that the generalized 
metric transforms as
\begin{equation}
\mathcal{H}(T,U) ~\stackrel{\hat\Sigma}{\longmapsto}~ \mathcal{H}(T',U') ~=~ \hat\Sigma^{-\mathrm{T}} \mathcal{H}(T,U)\, \hat\Sigma^{-1}\;.
\end{equation}
In this way, one can prove the transformations of the moduli, given in eqs.~\eqref{eq:SLZ2UandT} 
and~\eqref{eq:MandSigmaStar}, under the $\mathrm{O}_{\hat\eta}(2,2,\Z{})$ transformations listed in 
eqs.~\eqref{eq:ModularGroupOfT2Part1} and~\eqref{eq:ModularGroupOfT2Part2}.

For the (symmetric) $\mathbbm{T}^2/\Z{2}$ orbifold under consideration, the Narain twist is given 
by $\Theta = \hat\Theta = -\Id_4$. Hence, all outer automorphisms of the Narain lattice generated 
by the elements listed in eqs.~\eqref{eq:ModularGroupOfT2Part1} and~\eqref{eq:ModularGroupOfT2Part2} 
are also outer automorphisms of the $\Z{2}$ Narain space group. In addition, 
eq.~\eqref{eq:OuterOfSNarain} yields translational outer automorphisms 
$\hat{h}=\big(\Id_4,T\big)\not\in\hat{S}_\mathrm{Narain}$ that have to satisfy
\begin{equation}\label{eq:TranslationOnTwistedString}
\big(\hat\Theta, N\big) ~\stackrel{\hat{h}}{\longmapsto}~ \big(\Id_4,-T\big)\,\big(\hat\Theta, N\big)\,\big(\Id_4,T\big) ~=~ \big(\hat\Theta, N-(\Id_4-\hat\Theta)\,T\big) ~\stackrel{!}{\in}~ \hat{S}_\mathrm{Narain}\;.
\end{equation}
Thus, we obtain the condition $(\Id_4-\hat\Theta)\,T\in\Z{}^4$ on Narain translations with 
$T\notin\Z{}^4$. For $\hat\Theta=-\Id_4$ we find that the solutions of this condition can be 
generated by (cf.\ appendix~\ref{app:Lutowski})
\begin{equation}\label{eq:Ti}
T_1 ~=~ \frac{1}{2}\begin{pmatrix}1\\0\\0\\0\end{pmatrix} \;,\; T_2 ~=~ \frac{1}{2}\begin{pmatrix}0\\1\\0\\0\end{pmatrix}\;,\; T_3 ~=~ \frac{1}{2}\begin{pmatrix}0\\0\\1\\0\end{pmatrix}\quad\mathrm{and}\quad T_4 ~=~ \frac{1}{2}\begin{pmatrix}0\\0\\0\\1\end{pmatrix}\;.
\end{equation}

\subsection[Transformation of Z2 twisted strings]{\boldmath Transformation of $\Z{2}$ twisted strings\unboldmath}

From eq.~\eqref{eq:TranslationOnTwistedString} we find that Narain translations 
$\hat{h}_i=\big(\Id_4,T_i\big)\not\in\hat{S}_\mathrm{Narain}$ with $T_i$ given in eq.~\eqref{eq:Ti} 
act on the constructing elements $\big(\hat\Theta, N\big) \in \hat{S}_\mathrm{Narain}$ of $\Z{2}$ 
twisted matter fields $\phi_{(n_1,n_2)}$ as
\begin{equation}\label{eq:TOnTwistedStrings}
\big(\hat\Theta, N\big) ~\stackrel{\hat{h}_i}{\longmapsto}~ \big(\hat\Theta, N-2T_i\big) ~\in~\hat{S}_\mathrm{Narain}\;,
\end{equation}
using $\hat\Theta\,T_i = -T_i$. Note that winding numbers $(n_1,n_2)$ and KK numbers $(m_1,m_2)$ of 
\Z{2} twisted strings with $N=(n_1,n_2,m_1,m_2)^\mathrm{T}$ are defined modulo 2 (by considering 
the conjugacy classes $[\hat{g}]$ of constructing elements $\hat{g}\in \hat{S}_\mathrm{Narain}$). 
Consequently, eq.~\eqref{eq:TOnTwistedStrings} shows that $\hat{h}_1$ and $\hat{h}_2$ increase the 
winding number $n_1$ and $n_2$ by one unit, respectively. This confirms the geometrical intuition, 
illustrated in figure~\ref{fig:SpaceGroupAutomorphims}: $\hat{h}_1$ interchanges the twisted matter 
fields $\phi_{(0,n_2)}$ and $\phi_{(1,n_2)}$, while $\hat{h}_2$ interchanges $\phi_{(n_1,0)}$ and 
$\phi_{(n_1,1)}$. On the other hand, $\hat{h}_3$ and $\hat{h}_4$ act only on the KK numbers $m_1$ 
and $m_2$ of the twisted strings. Hence, each twisted matter field $\phi_{(n_1,n_2)}$ is mapped by 
$\hat{h}_3$ and $\hat{h}_4$ to itself, possibly times a phase. In summary, these considerations 
show that the Narain automorphism $\hat{h}_i=\big(\Id_4,T_i\big)$ gives rise to a representation 
$\rho_{\rep{4}}(h_i)$, for $i=1,\ldots,4$, as displayed in 
eqs.~\eqref{eq:OuterOfTwistedStrings},~\eqref{eq:SGn1} and~\eqref{eq:SGn2}.

\subsection{Details on mirror symmetry}
\label{app:mirror}

Let us consider the constructing elements $\hat{g}=\big(\hat\Theta, N\big)\in\hat{S}_\mathrm{Narain}$ 
of $\Z{2}$ twisted strings with $N=(n_1,n_2,m_1,m_2)^\mathrm{T}$ and $n_1,n_2,m_1,m_2\in\{0,1\}$. 
These 16 elements are associated with four twisted matter fields $\phi_{(n_1,n_2)}$ as follows:
\begin{equation}\label{eq:MappingPhiNarainN}
\phi_{(n_1,n_2)} \quad\leftrightarrow\quad \hat{g}~=~\big(\hat\Theta, N\big) \ \ \mathrm{with}\ \ N ~=~ (n_1,n_2,m_1,m_2)^\mathrm{T} \quad\mathrm{and}\quad m_1,m_2~\in~\{0,1\}\;.
\end{equation}
Then, according to eq.~\eqref{eq:OuterOfSNarainLatticeBasis} mirror symmetry $\hat{M}$ acts as
\begin{equation}\label{eq:MOnTwistedStrings}
\big(\hat\Theta, N\big) ~\stackrel{\hat{M}}{\longmapsto}~ \big(\hat\Theta, \hat{M}^{-1}\,N\big) ~\in~\hat{S}_\mathrm{Narain}\;,
\end{equation}
where we used $\hat{M}^{-1}\,\hat\Theta\,\hat{M} = \hat\Theta$ for the $\Z{2}$ Narain twist 
$\hat\Theta=-\Id_4$, and $\hat{M}$ is given in eq.~\eqref{eq:ModularGroupOfT2Part2}. Hence, we obtain
\begin{equation}\label{eq:MOnN}
\begin{pmatrix}n_1\\n_2\\m_1\\m_2\end{pmatrix} ~\stackrel{\hat{M}}{\longmapsto}~ \begin{pmatrix}0&0&-1&0\\0&1&0&0\\-1&0&0&0\\0&0&0&1\end{pmatrix}\,\begin{pmatrix}n_1\\n_2\\m_1\\m_2\end{pmatrix} ~=~ \begin{pmatrix}-m_1\\n_2\\ -n_1\\m_2\end{pmatrix}\;.
\end{equation}
Using the correspondence between constructing elements of the Narain space group and the twisted matter 
fields stated in eq.~\eqref{eq:MappingPhiNarainN}, we get  
\begin{subequations}
\begin{eqnarray}
\phi_{(0,0)} ~\stackrel{\hat{M}}{\longmapsto}~ \alpha_{11}\,\phi_{(0,0)} + \alpha_{12}\,\phi_{(1,0)}\;,\\
\phi_{(1,0)} ~\stackrel{\hat{M}}{\longmapsto}~ \alpha_{21}\,\phi_{(0,0)} + \alpha_{22}\,\phi_{(1,0)}\;,\\
\phi_{(0,1)} ~\stackrel{\hat{M}}{\longmapsto}~ \alpha_{33}\,\phi_{(0,1)} + \alpha_{34}\,\phi_{(1,1)}\;,\\
\phi_{(1,1)} ~\stackrel{\hat{M}}{\longmapsto}~ \alpha_{43}\,\phi_{(0,1)} + \alpha_{44}\,\phi_{(1,1)}\;,
\end{eqnarray}
\end{subequations}
with unknown coefficients $\alpha_{ij}$. Similar to eq.~\eqref{eq:ActionOnPhi}, this defines the 
matrix representation of the mirror transformation $\hat M$ on twisted matter fields
\begin{equation}
\tilde\rho_{\rep{4}}(\hat{M}) ~:=~ \begin{pmatrix}\alpha_{11}&\alpha_{12}&0&0\\ \alpha_{21}&\alpha_{22}&0&0\\0&0&\alpha_{33}&\alpha_{34}\\0&0&\alpha_{43}&\alpha_{44}\\ \end{pmatrix}\;,
\end{equation}
where we have used the notation $\tilde\rho_{\rep{4}}(\hat{M})$ (i.e.\ with a tilde) as it will
be redefined at the end of this section.

Moreover, one can analyze the group of outer automorphisms of $\hat{S}_\mathrm{Narain}$, especially 
concerning the translations $T_i$, $i=1,\ldots,4$, given in eq.~\eqref{eq:Ti} and mirror symmetry 
$\hat{M}$, i.e.
\begin{subequations}
\begin{eqnarray}
\big(\hat{M},0\big)\,\big(\Id_4,T_1\big)\,\big(\hat{M}^{-1},0\big) & = & \big(\Id_4,-T_3\big)\;,\\
\big(\hat{M},0\big)\,\big(\Id_4,T_2\big)\,\big(\hat{M}^{-1},0\big) & = & \big(\Id_4,T_2\big)\;,\\
\big(\hat{M},0\big)\,\big(\Id_4,T_3\big)\,\big(\hat{M}^{-1},0\big) & = & \big(\Id_4,-T_1\big)\;,\\
\big(\hat{M},0\big)\,\big(\Id_4,T_4\big)\,\big(\hat{M}^{-1},0\big) & = & \big(\Id_4,T_4\big)\;,
\end{eqnarray}
\end{subequations}
where $T_i$ are defined up to integers such that $-T_i \sim T_i$. Let us embed these 
equations into their action on twisted matter fields
\begin{subequations}
\begin{eqnarray}
\tilde\rho_{\rep{4}}(\hat{M})\,\rho_{\rep{4}}(h_1)\,\tilde\rho_{\rep{4}}(\hat{M})^{-1} & = & \rho_{\rep{4}}(h_3)\;,\\
\tilde\rho_{\rep{4}}(\hat{M})\,\rho_{\rep{4}}(h_2)\,\tilde\rho_{\rep{4}}(\hat{M})^{-1} & = & \rho_{\rep{4}}(h_2)\;,\\
\tilde\rho_{\rep{4}}(\hat{M})\,\rho_{\rep{4}}(h_3)\,\tilde\rho_{\rep{4}}(\hat{M})^{-1} & = & \rho_{\rep{4}}(h_1)\;,\\
\tilde\rho_{\rep{4}}(\hat{M})\,\rho_{\rep{4}}(h_4)\,\tilde\rho_{\rep{4}}(\hat{M})^{-1} & = & \rho_{\rep{4}}(h_4)\;.
\end{eqnarray}
\end{subequations}
This fixes all unknowns $\alpha_{ij}$ except for $\alpha_{11}$, 
\begin{equation}
\tilde\rho_{\rep{4}}(\hat{M}) ~=~ \alpha_{11}\,\begin{pmatrix}1&1&0&0\\ 1&-1&0&0\\0&0&1&1\\0&0&1&-1\\ \end{pmatrix}\;,
\end{equation}
and we obtain a representation of mirror transformation that is of order two by setting 
$\alpha_{11} = \nicefrac{1}{\sqrt{2}}$. However, in order to disentangle the two finite modular 
groups (associated with the K\"ahler modulus and the complex structure modulus) we have decided to 
redefine $\tilde\rho_{\rep{4}}(\hat{M})$ according to
\begin{equation}\label{eq:redef}
\rho_{\rep{4}}(\hat{M}) ~:=~ \tilde\rho_{\rep{4}}(\hat{M})\,\rho_{\rep{4}}(h_1)\,\rho_{\rep{4}}(h_2)\,\rho_{\rep{4}}(h_3)\,\rho_{\rep{4}}(h_4) ~=~ \frac{1}{\sqrt{2}}\begin{pmatrix}0&0&-1&1\\ 0&0&1&1\\1&-1&0&0\\-1&-1&0&0\\ \end{pmatrix}\;,
\end{equation}
as stated in eq.~\eqref{eq:repM} in section~\ref{sec:Modular}. This redefinition is possible, since 
each transformation $h_i$ associated with $\rho_{\rep{4}}(h_i)$ in eq.~\eqref{eq:redef} belongs to 
the traditional flavor symmetry that does not affect the moduli and, hence, is valid everywhere in 
moduli space. Further, the redefinition eq.~\eqref{eq:redef} does not alter the physics of the 
theory because $\rho_{\rep{4}}(h_i)$ is a symmetry transformation. Additional details will be given 
in ref.~\cite{Baur:2020pr}.

%%%%%%%%%%%%%%%%%%%%%%%%%%%%%%%%%%%%%%%%%%%%%%%%%%%%%%%%%%%%%%%%%%%%%%%%%%%%%%%%%%%%%%%%%%%%%%%%%%%%%%%%%%%%%%%%%%%%%%%%%%%%%%%%%%%%%%%%%%%%

\section{How to classify the outer automorphisms of a space group}
\label{app:Lutowski}

The group of outer automorphisms $\mathrm{Out}(\hat{S}_\mathrm{Narain})$ of a Narain space group 
$\hat{S}_\mathrm{Narain}$ is the key to uncover all discrete symmetries resulting from the orbifold 
compactification encoded in $\hat{S}_\mathrm{Narain}$. Interestingly, there exists a general 
algebraic construction of $\mathrm{Out}(S)$ for any kind of space group $S$ due to Lutowski, see 
refs.~\cite{lutowski2013finite,charlap1986bieberbach}. In this appendix, we briefly demonstrate its 
application to the Narain space group of the $\mathbbm{T}^2/\Z{2}$ orbifold in order to confirm the 
results presented in appendix~\ref{app:Out}.

The $\Z{2}$ Narain space group of our $\mathbbm{T}^2/\Z{2}$ orbifold is defined as 
\begin{equation}
\hat{S}_{\mathrm{Narain}} ~=~ \big\langle~ \big(\Id_{4}, N\big),\big(\hat\Theta, 0\big)~\big\vert ~ N \in \mathbbm{Z}^{4},~ \hat\Theta = - \Id_{4} ~\big\rangle 
\end{equation}
in the lattice basis. Then, Lutowski's algorithm states that $\mathrm{Out}(\hat{S}_\mathrm{Narain})$ 
is given by two factor groups, denoted by $\Xi$ and $\mathrm{H}^{1}(\hat{P},\mathbbm{Z}^{4})$. These 
groups combine semi-directly in the same way as the lattice and the point group combine to define 
$\hat{S}_\mathrm{Narain}$, i.e.\ 
\begin{equation}\label{eq:OutS}
\mathrm{Out}(\hat{S}_{\mathrm{Narain}}) ~=~ \Xi\ltimes\mathrm{H}^{1}(\hat{P},\mathbbm{Z}^{4})\;.
\end{equation}
The constituents of $\mathrm{Out}(\hat{S}_{\mathrm{Narain}})$ are the so-called {\it stabilizer $\Xi$
of the space group} and the {\it first cohomology group} $\mathrm{H}^{1}(\hat{P},\mathbbm{Z}^{4})$, 
where $\hat{P}$ is the Narain point group in the lattice basis 
($\hat{P}=\big\langle \hat\Theta \big\rangle=\{\pm\Id_4\}\cong\Z{2}$ in our case). Therefore, 
one can write an outer automorphism as $\hat{h}=\big(\hat{\Sigma},T\big)\not\in\hat{S}_\mathrm{Narain}$ 
with a rotational part $\hat{\Sigma}$ and a translational part $T$ that acts as
\begin{equation}
\big(\hat\Theta^k, N\big) ~\stackrel{\hat{h}}{\longmapsto}~ \big(\hat{\Sigma},T\big)^{-1}\,\big(\hat\Theta^k, N\big)\,\big(\hat{\Sigma},T\big) ~\in~ \hat{S}_{\mathrm{Narain}}
\end{equation}
for all $\big(\hat\Theta^k, N\big)\in\hat{S}_{\mathrm{Narain}}$. By applying Lutowski's algorithm, 
one observes that $\mathrm{H}^{1}(\hat{P},\mathbbm{Z}^{4})$ accounts for pure translations. In 
detail, one finds
\begin{equation}
\mathrm{H}^{1}(\hat{P},\mathbbm{Z}^{4}) ~=~ \big\langle ~T_i~ \big\vert ~ i=1,2,3,4  ~\big\rangle ~/~ \mathbbm{Z}^{4}~\cong~ \left( \Z{2} \right)^{4}\;,
\end{equation}
see eq.~\eqref{eq:Ti}. The four generators of this group $T_i$ correspond exactly to the 
transformations $\hat{h}_{i}$ introduced earlier as the geometrical translations for $i=1,2$ and 
the space group selection rule for $i=3,4$. Acting on the twisted strings of the $\Z{2}$ orbifold, 
these outer automorphisms give rise to the traditional flavor symmetry.

The other factor of $\mathrm{Out}(\hat{S}_{\mathrm{Narain}})$ in eq.~\eqref{eq:OutS} is the 
stabilizer $\Xi$, which is governed by the explicit form of the space group. In general, an element 
of $\Xi$ might admit both, a rotational part $\hat\Sigma$ and a translational part $s(\hat\Sigma)$ 
such that a general element is given by a so-called roto-translation 
$\big(\hat\Sigma, s(\hat\Sigma)\big)$. Since the $\Z{2}$ point group 
$\hat{P}=\big\langle \hat\Theta \big\rangle$ considered here acts as a pure rotation on the Narain 
lattice, we can set $s(\hat\Sigma)=0$ and the algorithm finds the stabilizer group to be 
\begin{equation}
\mathrm{GL}(4,\mathbbm{Z})~/~\big\langle \hat\Theta \big\rangle\;.
\end{equation}
However, we have to impose the physical condition (related to level-matching) that the rotation 
$\hat\Sigma$ has to preserve the Narain metric in the lattice basis $\hat\eta$, given in 
eq.~\eqref{eq:NarainMetric}. This means that $\hat\Sigma^{\mathrm{T}}\hat\eta\hat\Sigma=\hat\eta$ 
has to hold. As defined in eq.~\eqref{eq:OEtaHat22Z}, these rotations form the subgroup 
$\mathrm{O}_{\hat\eta}(2,2, \mathbbm{Z})$ of $\mathrm{GL}(4,\mathbbm{Z})$. Therefore, the 
stabilizer group $\Xi$ of the $\Z{2}$ Narain space group is given by
\begin{equation}
\label{eq:LutowskiStabilizer}
\Xi ~\cong~ \mathrm{O}_{\hat\eta}(2,2,\mathbbm{Z})~/~\big\langle \hat\Theta \big\rangle\;.
\end{equation}
Its set of elements is generated by $\hat{K}_{\mathrm{S}}$, $\hat{K}_{\mathrm{T}}$, 
$\hat{C}_{\mathrm{S}}$, $\hat{C}_{\mathrm{T}}$, $\hat{\Sigma}_{*}$ and $\hat{M}$, as listed in 
eqs.~\eqref{eq:ModularGroupOfT2Part1} and~\eqref{eq:ModularGroupOfT2Part2}, modulo point group 
transformations with $\hat\Theta=-\Id_4$. This class of outer automorphisms of the Narain space 
group generates the group of modular transformations of our $\mathbbm{T}^2/\Z{2}$ compactification 
background.

Finally, with the semi-direct product being mediated by the Narain version of the group law in 
eq.~\eqref{eq:SpaceGroupGroupLaw}, the outer automorphisms of the Narain space group reads
\begin{equation}
\mathrm{Out}(\hat{S}_{\mathrm{Narain}}) = \big\langle~ \big( \hat\Sigma, 0\big),\big( \Id_{4} , T_i\big) ~\vert ~ \hat\Sigma\in  \big\lbrace \hat{K}_{\mathrm{S}},\hat{K}_{\mathrm{T}},\hat{C}_{\mathrm{S}},\hat{C}_{\mathrm{T}},\hat{\Sigma}_{*},\hat{M} \big\rbrace,~i=1,\ldots,4 ~\big\rangle ~/~ \hat{S}_{\mathrm{Narain}}\;.
\end{equation}

%%%%%%%%%%%%%%%%%%%%%%%%%%%%%%%%%%%%%%%%%%%%%%%%%%%%%%%%%%%%%%%%%%%%%%%%%%%%%%%%%%%%%%%%%%%%%%%%%%%%%%%%%%%%%%%%%%%%%%%%%%%%%%%%%%%%%%%%%%%%

\section{Irreducible representations of twisted matter fields}
\label{app:irreps}

The four twisted matter fields $(\phi_{(0,0)}, \phi_{(1,0)}, \phi_{(0,1)}, \phi_{(1,1)})^\mathrm{T}$ 
are localized at the four fixed points of the $\mathbbm T^2/\Z2$ orbifold. They transform under 
both, the traditional flavor symmetry $(D_8\x D_8)/\Z2\cong[32,49]$, where they transform as an 
irreducible $\rep{4}$, and also under the modular symmetries discussed in section~\ref{sec:Modular}. 
There, it was found that twisted matter fields do not transform faithfully under $\SL{2,\Z{}}_T$ 
and $\SL{2,\Z{}}_U$ but in a four-dimensional unitary representation $\rho_{\rep{4}}$ of the 
respective finite modular groups $S_3^T$ and $S_3^U$. Moreover, this representation turns out to be 
reducible. In detail, by studying the characters of the representation matrices one can show that 
the four-dimensional representation decomposes i) into irreducible representations of each $S_3$ 
factor as
\begin{equation}
\rep{4} ~=~ \rep{2} \oplus \rep{1} \oplus \rep{1}
\end{equation}
and ii) into irreducible representations of $S_3^T \times S_3^U$ as
\begin{equation}
\rep{4} ~=~ (\rep{2},\rep{1}) \oplus (\rep{1},\rep{2})\;.
\end{equation}
Moreover, including mirror symmetry, $S_3^T \times S_3^U$ is enhanced to $[144,115]$ (cf.\ 
eq.~\eqref{eq:GenericFiniteModularGroupNoCP}), where the four twisted matter fields build an 
irreducible representation $\rep{4}$. These decompositions can be made explicit by the following 
orthogonal basis change $B$:
\begin{equation}
\begin{pmatrix}\phi_1\\\phi_2\\\phi_3\\\phi_4\end{pmatrix}
~:=~
B\,
\begin{pmatrix}
\phi_{(0,0)}\\ \phi_{(1,0)}\\ \phi_{(0,1)}\\ \phi_{(1,1)}
\end{pmatrix}\;,\quad\mathrm{where}\quad
B ~:=~ \begin{pmatrix}
0 & \nicefrac{-1}{\sqrt{3}} & \nicefrac{-1}{\sqrt{3}} & \nicefrac{-1}{\sqrt{3}} \\
1 &                       0 &                       0 & 0 \\
0 & \nicefrac{-2}{\sqrt{6}} &  \nicefrac{1}{\sqrt{6}} & \nicefrac{1}{\sqrt{6}} \\
0 &                       0 &  \nicefrac{1}{\sqrt{2}} & \nicefrac{-1}{\sqrt{2}} 
\end{pmatrix}\;.
\end{equation}
In this basis, the matrix representations, 
eqs.\ (\ref{eq:SL2ZUOfTwistedStrings})-(\ref{eq:SL2ZTOfTwistedStrings}), of modular 
transformations of the four twisted matter fields $\phi_n$ are given by 
$\rho_{\rep{4}}'(\gamma) = B\,\rho_{\rep{4}}(\gamma)\,B^{-1}$. They take the following form
\begin{subequations}
\begin{align}
&\rho_{\rep{4}}'(\hat{K}_{\mathrm{S}}) ~=~ 
\left(\begin{array}{cc|cc}
\nicefrac{-1}{2} & \nicefrac{-\sqrt{3}}{2} & 0 & 0 \\
\nicefrac{-\sqrt{3}}{2} & \nicefrac{1}{2} & 0 & 0 \\\hline 
0 & 0 & 1 & 0\\
0 & 0 & 0 & 1
\end{array}\right)\;,
&&
\rho_{\rep{4}}'(\hat{K}_{\mathrm{T}}) ~=~ 
\left(\begin{array}{cc|cc}
1 & 0 & 0 & 0 \\
0 & -1 & 0 & 0 \\\hline 
0 & 0 & 1 & 0\\
0 & 0 & 0 & 1
\end{array}\right)\;,\\[6pt]
&\rho_{\rep{4}}'(\hat{C}_{\mathrm{S}}) ~=~ 
\left(\begin{array}{cc|cc}
1 & 0 & 0 & 0 \\
0 & 1 & 0 & 0 \\\hline 
0 & 0 & \nicefrac{-1}{2} & \nicefrac{-\sqrt{3}}{2} \\
0 & 0 & \nicefrac{-\sqrt{3}}{2} & \nicefrac{1}{2}
\end{array}\right)\;,
&&
\rho_{\rep{4}}'(\hat{C}_{\mathrm{T}}) ~=~ 
\left(\begin{array}{cc|cc}
1 & 0 & 0 & 0 \\
0 & 1 & 0 & 0 \\\hline 
0 & 0 & 1 & 0\\
0 & 0 & 0 & -1
\end{array}\right)\;,
\end{align} \\
where we indicate the $2\times 2$ block structure by horizontal and vertical lines, cf. 
ref.~\cite{Ishimori:2010au}. This proves the 
$(\rep{2},\rep{1}) \oplus (\rep{1},\rep{2})$ block structure with respect to $S_3^T\times S_3^U$. 
Furthermore, in this basis it is easy to see that the mirror transformation $\hat{M}$ interchanges 
$S_3^T$ and $S_3^U$, i.e.
\begin{equation}
\rho_{\rep{4}}'(\hat{M}) ~=~ 
\left(\begin{array}{cc|cc}
0 & 0 & -1 & 0 \\
0 & 0 & 0 & -1 \\\hline 
1 & 0 & 0 & 0\\
0 & 1 & 0 & 0
\end{array}\right)\;.
\end{equation} 
\end{subequations}
Finally, the irreducible representations in which the twisted matter fields transform under 
various components of the eclectic flavor symmetry are summarized in table~\ref{tab:Irreps}.

\begin{table}
\centering
\begin{tabular}{cccc|c}
\toprule
\multicolumn{4}{c|}{finite modular symmetry} & traditional flavor symmetry \\
$S_3^T$                            & $S_3^U$                            & $S_3^T\times S_3^U$                        & $(S_3^T \times S_3^U)\rtimes \Z{4}^{\hat{M}}$ & $(D_8 \times D_8)/\Z{2}$ \\\midrule
$\rep{2}\oplus\rep{1}\oplus\rep{1}$& $\rep{2}\oplus\rep{1}\oplus\rep{1}$& $(\rep{2},\rep{1})\oplus(\rep{1},\rep{2})$ & $\rep{4}$                            & $\rep{4}$ \\ 
\bottomrule
\end{tabular}
\caption{Irreducible representations of twisted matter fields $(\phi_{(0,0)}, \phi_{(1,0)}, \phi_{(0,1)}, \phi_{(1,1)})^\mathrm{T}$ 
with respect to the various flavor symmetries (in the absence of string oscillator excitations).}
\label{tab:Irreps}
\end{table}

%\bibliography{Orbifold}
%\bibliographystyle{OurBibTeX}
\providecommand{\bysame}{\leavevmode\hbox to3em{\hrulefill}\thinspace}

\end{document}